\documentclass[10pt,journal]{IEEEtran}
\IEEEoverridecommandlockouts
\usepackage{cite}
\usepackage{epsfig}
\usepackage{tikz}
\usepackage{lettrine}
\usepackage{enumitem}
\usepackage{dirtytalk}
\usepackage[table]{xcolor}
\usepackage{amsthm}

\newtheorem{lemma}{Lemma}
\usepackage{amsmath,amssymb,amsfonts}
\allowdisplaybreaks
\usepackage{mathrsfs}
\usepackage{algorithmic}
\usepackage{graphicx}
\usepackage{subcaption}
\usepackage{textcomp}
\usepackage{xcolor}
\usepackage{amsmath}
\usepackage{cuted}
\usepackage{xparse}
\usepackage{amsfonts}
\usepackage[multiple]{footmisc}
\usepackage{amssymb}
\usepackage{mathtools}
\usepackage{stackengine}
\usepackage[mathscr]{euscript}
\usepackage{graphics}
\usepackage[T1]{fontenc}
\usepackage{verbatim} 
\usepackage{multicol}
\usepackage{svg}
\usepackage{lettrine}
\usepackage[table]{xcolor}
\usepackage{etoolbox}
\allowdisplaybreaks
\usepackage{mathrsfs}
\usepackage{afterpage}
\usepackage[font=footnotesize,labelfont=footnotesize]{caption}
\usepackage[ruled,vlined]{algorithm2e}
\usepackage{listings}
\usepackage{url}
\usepackage{pdfpages}
\usepackage{float}
\usepackage{mathptmx}
\usepackage[11pt]{moresize}
\usepackage{stfloats}
\usepackage{diagbox}
\usepackage{wrapfig}
\usepackage{balance}
\allowdisplaybreaks
\usepackage[caption=false,font=footnotesize]{subfig}
\makeatother
\title{Secrecy and Covertness Analysis of SWAN-Enabled Uplink EDMA-NOMA Systems \vspace{-0.25em}}

\author{Sagnik Bhattacharyya, ~\IEEEmembership{Member, IEEE}, Keshav Singh, ~\IEEEmembership{Senior Member, IEEE},\\ Shiv Kumar, ~\IEEEmembership{Member, IEEE}, and Chih-Peng Li,~\IEEEmembership{Fellow,~IEEE}

\thanks{Sagnik Bhattacharyya, Keshav Singh, and Chih-Peng Li are with the Institute of Communications Engineering, National Sun Yat-sen University, Kaohsiung 80424, Taiwan (e-mail: sagnikbhattacharyya95@gmail.com, keshav.singh@mail.nsysu.edu.tw, cpli@faculty.nsysu.edu.tw).}

\thanks{Shiv Kumar is with the Department of Electronics and Communication Engineering, Netaji Subhas University of Technology, New Delhi, 110078, India (e-mail: shiv@nsut.ac.in).}

\vspace{-2.5em}
}
\begin{document}
\maketitle
\begin{abstract}
   Environment division multiple access (EDMA) has recently emerged as a promising multiple-access paradigm by exploiting the wireless propagation environment for user separation, while physical-layer security, including secrecy and covert communication, is becoming increasingly important for protecting confidential information and user privacy. Motivated by these considerations, this paper proposes an EDMA-enabled non-orthogonal multiple access for secrecy and covert communications (EDMA-NOMA-SC) framework for uplink wireless systems. In the proposed framework, a segmented-waveguide-enabled pinching antenna system (SWAN) is employed, where a secrecy user and an intermittently active covert user share the same transmission resource through NOMA within each segment, while a malicious warden attempts to eavesdrop on the secrecy transmission and detect the presence of the covert communication. To capture practical operating conditions, distance-dependent blockage, inter-segment interference, in-waveguide attenuation, imperfect channel state information (ipCSI), and imperfect successive interference cancellation (ipSIC) are considered. Semi-analytical frameworks are developed to characterize the secrecy outage probability (SOP), average secrecy capacity (ASC), false alarm probability (FAP), missed detection probability (MDP), detection error probability (DEP), and covert rate (CR), with their accuracy verified through Monte Carlo simulations. Numerical results, including comparisons with conventional and EDMA-based benchmark schemes, show that the proposed framework generally achieves improved secrecy and covert-rate performance owing to reduced user-to-pinching-antenna propagation distances and in-waveguide attenuation. However, excessive segmentation can degrade performance when inter-segment interference outweighs these propagation benefits. The results further show that ipCSI and ipSIC degrade the achievable performance, while inter-segment interference can increase Willie’s detection uncertainty and thereby improve covertness under suitable operating conditions.
\end{abstract}
\vspace{-0.5em}
\vspace{-1.5em}
\section{Introduction}
\IEEEPARstart The rapid evolution of wireless communication technologies has transformed wireless networks into a key infrastructure for increasingly intelligent, interconnected, and mission-critical applications. With the global deployment of fifth-generation (5G) networks, substantial advances have been achieved in data rate, latency, connectivity, reliability, and energy efficiency \cite{Habibi}. Nevertheless, sixth-generation (6G) wireless networks are expected to support even higher levels of connectivity and service heterogeneity, accompanied by the exchange of massive volumes of sensitive and mission-critical information \cite{dang2020should}. Owing to the broadcast nature of wireless propagation, such pervasive connectivity also increases the exposure of transmitted information to unauthorized entities, making security and privacy fundamental requirements for future wireless systems \cite{VanNguyen,Xiqi}. Traditionally, wireless security has primarily relied on cryptographic techniques implemented at the upper layers of the communication protocol stack. However, their key-management requirements and computational complexity can be challenging for highly dynamic and resource-constrained wireless networks. In this context, physical-layer security (PLS) has emerged as an effective complementary approach by exploiting the inherent randomness and spatial characteristics of wireless channels \cite{VanNguyen}. In particular, information-theoretic secrecy aims to ensure reliable information delivery to legitimate users while restricting information leakage to potential eavesdroppers \cite{Scott}. However, secrecy alone does not conceal the existence of a wireless transmission, since an adversary may fail to decode the confidential message while still detecting its presence. This motivates covert communication, which aims to prevent a vigilant warden from reliably distinguishing between the absence and presence of a transmission \cite{Chen}. Thus, secrecy and covertness address complementary security requirements of protecting confidential information and concealing the communication itself. Motivated by these security requirements, recent studies have also investigated uplink non-orthogonal multiple access (NOMA) frameworks with heterogeneous security demands, including joint secrecy and covert transmission \cite{10938035} and covert communications in the presence of a warden \cite{10459011}. Since both strongly depend on the relative channel conditions of legitimate and adversarial users, additional spatial degrees of freedom for controlling these channels are highly desirable. Accordingly, multi-antenna and multiple-input multiple-output (MIMO) systems have been widely investigated for secrecy and covert communications, where beamforming and related spatial-processing techniques strengthen legitimate links while suppressing information leakage and transmission detectability \cite{11552029,11052758}. \textit{However, conventional fixed-position antenna architectures have limited flexibility in fully exploiting spatial-diversity gains, while increasing the number of antenna elements to obtain additional spatial degrees of freedom can also increase hardware cost and implementation complexity.}
\par To overcome these limitations of conventional fixed-position antenna architectures, several reconfigurable wireless technologies, including reconfigurable intelligent surfaces (RISs) \cite{Swindlehurst}, fluid antenna systems (FASs) \cite{Farshad}, and movable antennas (MAs) \cite{Jingze}, have been investigated to provide additional flexibility in shaping wireless propagation. In particular, RISs reconfigure the propagation environment through adjustable reflecting elements, whereas FASs and MAs exploit flexible antenna positioning to obtain additional spatial degrees of freedom. These capabilities have been increasingly explored for PLS, where reconfigurable propagation and antenna positioning can strengthen legitimate links while suppressing information leakage toward potential eavesdroppers \cite{Swindlehurst,Farshad,Jingze}. Similarly, RIS-, FAS-, and MA-assisted architectures have also been investigated for covert communications \cite{Kaikit,11208163,11551156,11498624}. \textit{Nevertheless, FAs and MAs are generally restricted to movements over only a few wavelengths, resulting in limited control over large-scale path loss \cite{TCOM}. Moreover, the number of active antennas is typically fixed after deployment, which can restrict the available spatial degrees of freedom and thereby limit the achievable secrecy and covertness performance.}
\par Recently, pinching antenna systems (PASS) have emerged as a promising architecture for flexible antenna deployment \cite{10945421,TCOM,WCL,Tall2612:Sum}. In PASS, radio frequency signals propagate through one or multiple dielectric waveguides, where small pinching antennas (PAs) are attached to radiate the guided signals into free space. By dynamically adjusting the PA locations, PASS can establish favourable line-of-sight (LoS) links, reduce the free-space propagation distance, and manipulate the effective wireless channels over a much larger spatial scale than conventional fixed-position antenna architectures. Owing to these advantages, PASS has attracted considerable interest for PLS \cite{Naofal,11205176,papanikolaou2025physical,11576107}. Existing studies have exploited PA positioning and pinching beamforming to enhance the legitimate-user channel while suppressing information leakage toward eavesdroppers, with extensions considering artificial-noise transmission, multiple-waveguide architectures, secure multicast transmission, and secrecy-performance analysis. Similarly, PASS has recently been investigated for covert communications, where the controllable PA positions can reshape the relative channels of legitimate users and wardens, improve the covert rate, and increase detection uncertainty \cite{YuanweiJSACPASS,11426943,11613999,Shiv,11494673}. Recent works have considered PA-position optimization, sensing-assisted warden tracking, NOMA-based covert communication, UAV-assisted covert transmission, near-field covert communication, and covert backscatter systems. \textit{However, conventional PASS employing long waveguides can suffer from considerable in-waveguide attenuation owing to the larger propagation distance between the PA and its corresponding feed point (FP).} To address this limitation, segmented waveguide-enabled pinching-antenna systems (SWANs) have been proposed, where long waveguides are divided into multiple shorter segments, thereby reducing the PA-to-feed propagation distance and the associated in-waveguide attenuation \cite{11348983}. \textit{Despite these advantages, existing SWAN studies have primarily focused on conventional communication performance, while their potential for PLS, particularly for secrecy and covert communications, remains unexplored.}
\par Beyond wireless channel reconfiguration, flexible-antenna architectures also open new possibilities for multiple access schemes. Conventional multiple-access schemes distinguish users through the time, frequency, code, spatial, and power domains, which can restrict the resources available to individual users and limit spectral utilization. This motivates the exploration of additional access dimensions beyond conventional radio-resource partitioning. In this context, environment division multiple access (EDMA) has recently been developed to exploit propagation characteristics such as LoS blockage and large-scale path loss for suppressing co-channel interference and facilitating efficient multi-user transmission \cite{ding2025environment}. Now, owing to its advantages, PASS provides a suitable platform for EDMA by strengthening desired links while weakening interfering ones. SWAN further facilitates EDMA by dividing the service area into multiple spatial regions served through independently fed waveguide segments. However, existing SWAN-enabled EDMA studies mainly consider a single user per segment. More recently, EDMA has been integrated with NOMA to support multiple users within each segment in a downlink scenario \cite{Bhat2612:Performance}. \textit{Nevertheless, EDMA-NOMA for PLS, particularly joint secrecy and covert communications in the uplink, remains unexplored.} 
\par In this context, the integration of EDMA, NOMA, and SWAN provides a promising framework for wireless networks, where communication security and privacy are becoming increasingly crucial. Accordingly, investigating PLS. Motivated by these considerations, we propose an EDMA-enabled NOMA for secrecy and covert communications (EDMA-NOMA-SC) framework for uplink wireless systems, where a SWAN architecture partitions the service area into multiple segments and, within each segment, a secrecy user and an intermittently active covert user share the same transmission resources through NOMA in the presence of a warden. The \textbf{main contributions} of this paper are summarized as follows
\begin{itemize}
    \item \textbf{Introduction to EDMA-NOMA-SC framework:} We propose an EDMA-NOMA-SC framework for uplink wireless systems. The proposed framework employs a SWAN architecture and partitions the overall service area into multiple segments, where a secrecy user and an intermittently active covert user share the same transmission resource through NOMA within each segment, while a malicious warden attempts to eavesdrop on the secrecy transmission and detect the presence of the covert communication. The framework accounts for distance-dependent blockage, inter-segment interference, in-waveguide attenuation, imperfect channel state information (ipCSI), and imperfect successive interference cancellation (ipSIC). Specifically, ipCSI is characterised using a pilot-symbol-assisted channel-estimation model, wherein the estimation-error variance varies with the pilot SNR and average channel power. Based on the proposed architecture, the received signal models for the secrecy and covert communications are formulated, followed by the corresponding signal-to-noise ratio (SNR), signal-to-interference-plus-noise ratio (SINR), and received signal-energy expressions required for the performance analyses.
    \item \textbf{Secrecy Performance Analysis:} Ssemi-analytical expressions for the secrecy outage probability (SOP) and average secrecy capacity (ASC) under both covert-user activity hypotheses are derived. A direct characterization is analytically intractable because the secrecy event jointly depends on multiple random components, including user locations, blockage states, inter-segment non-line-of-sight (NLoS) fading, adjacent-segment covert-user activity, inter-segment interference, and intra-segment fading channels, resulting in a large number of possible network configurations. To make the analysis tractable, we condition on the random network configuration and analytically average over the remaining intra-segment channel randomness. The resulting integrals are evaluated using numerical integration methods and subsequently averaged over independent network realizations, thereby capturing the effects of ipCSI, ipSIC, inter-segment interference, and covert-user activity on the secrecy performance.
    \item \textbf{Covert Communication Performance Analysis:} Owing to the large number of possible network realizations arising from the random network parameters discussed above, obtaining compact closed-form expressions for the covert performance is analytically intractable. Therefore, we adopt a semi-analytical approach in which the aggregate inter-segment interference is characterized over independent network realizations to derive the false alarm probability (FAP), missed detection probability (MDP), and detection error probability (DEP). For the covert rate (CR), the remaining expectation over the intra-segment channel powers is evaluated using the composite midpoint quadrature rule.
    \item \textbf{Performance Comparison:} The proposed framework is compared with EDMA-OMA, conventional NOMA-PASS, and conventional OMA-PASS benchmark schemes under ideal and practical conditions, with the developed semi-analytical results validated through Monte Carlo simulations. For secrecy, EDMA-NOMA-SC generally provides improved SOP and ASC owing to reduced propagation distance and in-waveguide attenuation, although excessive segmentation can reverse this advantage when inter-segment interference becomes dominant. For covertness, inter-segment interference can increase Willie’s detection uncertainty depending on the operating conditions, while the proposed framework generally provides improved CR performance.
\end{itemize}
\begin{table}[t]
\centering
\caption{List of Notations and Mathematical Operators.}
\label{Notation}
\begin{tabular}{|>{\centering\arraybackslash}m{1.6cm}||p{6cm}|}
 \hline
 \rowcolor{magenta!20}\textbf{Symbols} &\hspace{20mm} \textbf{Description}\\
 \hline
  \rowcolor{cyan!20}\multicolumn{2}{|c|}{\textbf{Indices, Nodes, and Sets}} \\
  \hline
  $i,n$ & $\in\{1,\ldots,N\}$\\
  $Q$ & $\in\{B,C,W\}$: Ground nodes (Bob/Carol/Willie) in a segment\\
  $R$ & $\in\{B,C\}$: Transmitting nodes in a segment\\
  $X$ & $\in\{P,W\}$: Receiving/Eavesdropping nodes in a segment\\
  $q$ & $\in\{0,1\}$: Hypothesis index\\
  \hline
  \rowcolor{cyan!20}\multicolumn{2}{|c|}{\textbf{General System Parameters}} \\
  \hline
  $N, \EuScript{S}_n, \EuScript{A}_n$ & Total number of segments, the $n^{th}$ segment, and set of segments adjacent to $\EuScript{S}_n$, respectively\\
  $L, W$ & Length and width of the total area considered\\
 $d$ & Waveguide segment height\\
 $U_{R,n}, U_{X,n}$ & Bob (or Carol), Alice (or Willie) of $\EuScript{S}_n$, respectively\\
 $\sigma_{P,n}^2, \sigma_{W,n}^2$ & AWGN noise at $U_{P,n}$ and $U_{W,n}$, respectively\\
 $\EuScript{P}_{R,n}$ & Transmit power of $U_{R,n}$\\
 $\EuScript{G}$ & Random network configuration\\
 \hline
  \rowcolor{cyan!20}\multicolumn{2}{|c|}{\textbf{Channel Parameters}} \\
  \hline
 $h_{\omega,n}$ & In-waveguide channel of $\EuScript{S}_n$\\
 $h_{R,i}^{X,n}$ & Wireless channel between $U_{R,i}$ and $U_{P,n}$ (or $U_{W,n}$)\\ 
 $\eta_{\text{eff}}$ & Effective refractive index of any segment\\
 $\kappa$ & In-waveguide attenuation coefficient\\
 $\nu$ & NLoS path-loss exponent\\
 $c$, $f_C$ & Speed of light and carrier frequency, respectively\\
 $\Phi_{R,n}, \Phi_{X,n}$ & Coordinates of $U_{R,n}$ and $U_{P,n}$, respectively\\
 $\EuScript{N}_{R,n}$ & Number of pilot symbols\\
 $\rho_{R,n}$ & Channel estimation correlation coefficient\\
 $\theta_{e,R,n}^2$ & Channel-estimation error variance\\
 $\Delta_{P,n}$ & Residual interference coefficient due to ipSIC\\
 $\phi_{R,i}^{X,n}$ & Distance dependent channel blockage parameter\\
  \hline
  \rowcolor{cyan!20}\multicolumn{2}{|c|}{\textbf{Activity and Link Availability Indicators}} \\
  \hline
  $\alpha_{R,i}^{X,n}$ & Link availability indicator between $U_{R,i}$ and $U_{X,n}$\\
  $\beta_{C,n}$ & $\in\{0,1\}$: Carol activity indicator in $\EuScript{S}_n$\\
  $p_{act}$ & Probability of covert user being active\\
 \hline
 \rowcolor{cyan!20}\multicolumn{2}{|c|}{\textbf{Mathematical Operators/ Symbols}} \\
 \hline
 $j$, $[.]^+$, $\lVert . \rVert$, $\lvert . \rvert$, $\mathbb{E}[.]$ & Imaginary unit $\left(\sqrt{-1}\right)$, $\max(.,0)$, Euclidean norm, absolute value and mean operators, respectively\\
 $\Re\{.\}$, $\Im\{.\}$ & Real and imaginary part of a complex signal/ stream, respectively\\
 $\EuScript{U}[.,.], \EuScript{N}\{.\}$& Uniform and Normal distribution, respectively\\
 $\EuScript{CN}\{.,.\}$ & Complex normal distribution\\
 \hline
\end{tabular}
\end{table}
\par \textit{\textbf{Notations Used:}} The notation and mathematical operators used in the work are listed in Table \ref{Notation}.
\par \textit{\textbf{Organization of the paper:}} The remainder of the paper is organised as follows. Section \ref{systemmodel} presents the system model along with the corresponding signal modelling. Section \ref{PerformanceAnalysis} provides the SOP, DEP, ASC and CR analysis. Section \ref{Results} presents the analytical and simulation results, and Section \ref{Conclusion} finally concludes the paper.
\section{System Model}
\label{systemmodel}
\begin{figure*}[ht]
    \centering
    \includegraphics[scale=0.23]{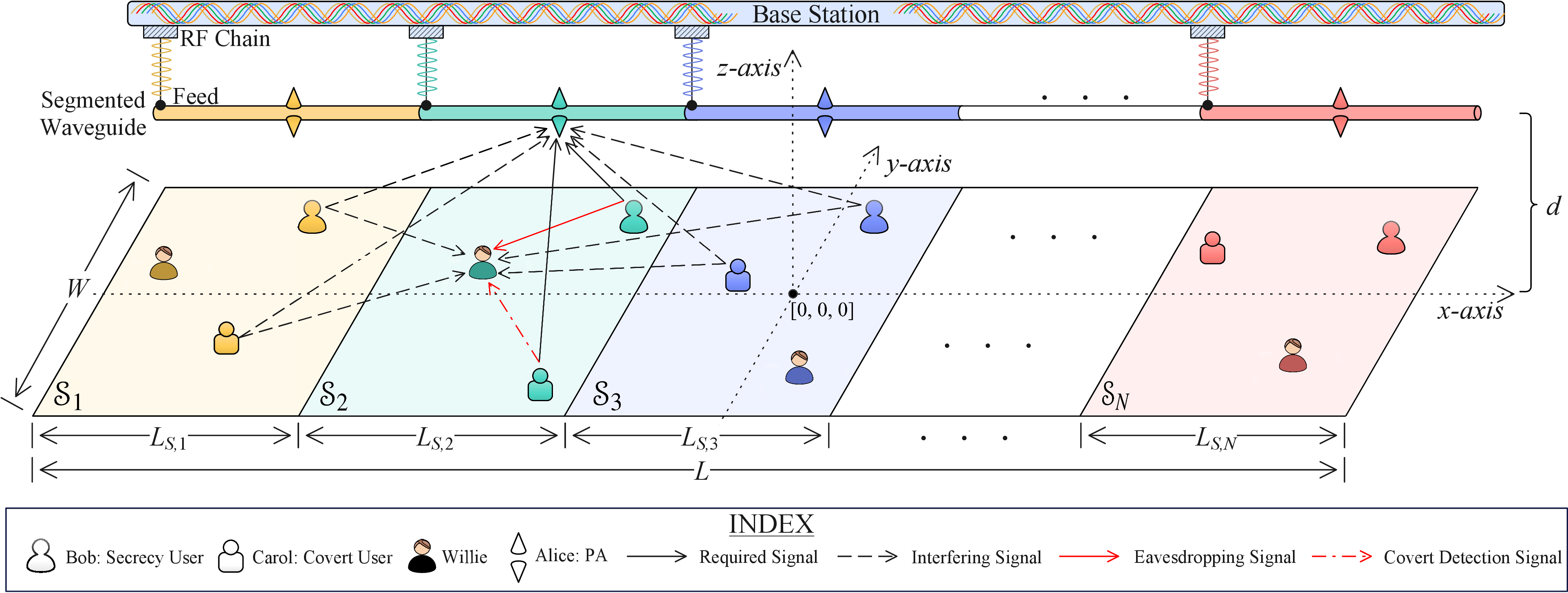}
    \caption{An illustration of the proposed uplink SWAN-enabled EDMA framework with NOMA-based secrecy and covert communications.}
    \label{GeneralDiagram}
\end{figure*}

Let us consider an uplink SWAN-enabled EDMA system, as illustrated in \figurename{~\ref{GeneralDiagram}}, serving a rectangular region of length $L$ and width $W$. The service region is partitioned into $N$ non-overlapping segments, denoted by $\EuScript{S}_n\forall n\in\{1,\ldots,N\}$. For simplicity, all segments are assumed to have identical dimensions, with segment length $L_S=L_{S,n}=\frac{L}{N}$ and width $W$. The segmented dielectric waveguide is deployed at a height $d$ above the service region, where each waveguide segment serves a corresponding spatial region. Each waveguide segment is connected to the base station (BS) through a dedicated RF chain and an individual feed point located at $\Phi_{F,n}=[x_{F,n},0,d]$, where $x_{F,n}=\frac{-L}{2}+(n-1)L_S$. Moreover, each waveguide segment is equipped with a dedicated PA (also referred to as Alice). For the $n^{th}$ segment, the corresponding PA is denoted by $U_{P,n}$, whose position within $\EuScript{S}_n$ is given by $\Phi_{P,n}=[x_{P,n},0,d]$, where $x_{P,n}=\frac{-L}{2}+\left(n-\frac{1}{2}\right)L_S$. This placement provides a common and symmetric reference configuration across the segments and facilitates the subsequent performance analysis. A similar fixed PA configuration has also been adopted in \cite{Bhat2612:Performance}. The overall SWAN configuration therefore follows the \textit{segment-multiplexing} (SM) principle \cite{11348983}.
\par Each spatial region served by a corresponding waveguide segment accommodates two legitimate ground users, Bob and Carol. In $\EuScript{S}_n$, these users are denoted by $U_{B,n}$ and $U_{C,n}$, respectively. Both users transmit their signals to the PA associated with the $n^{th}$ segment. In addition, each spatial region contains a malicious node, Willie\footnote{In this work, the Willies across segments are assumed to be non-colluding. The analysis can be extended in future work to a colluding-Willie scenario, where malicious nodes located in different segments cooperate by sharing their received observations of signals originating from adjacent segments.}, denoted by $U_{W,n}$, which attempts to eavesdrop on Bob’s confidential transmission while detecting the presence of Carol’s communication. Accordingly, Bob requires secrecy protection for his transmitted information, whereas Carol requires covert transmission to conceal the occurrence of his communication. The Cartesian coordinate of any such node $U_{Q,n}\forall Q\in\{B,C,W\}$, is denoted by $\Phi_{Q,n}=\left[x_{Q,n},y_{Q,n},0\right]$ where $x_{Q,n}\sim\EuScript{U}\left[\frac{-L}{2}+(n-1)L_S,\frac{-L}{2}+nL_S\right]$, and $y_{Q,n}\sim\EuScript{U}\left[\frac{-W}{2},\frac{W}{2}\right]$. This setup defines the considered network geometry, including the locations of the users and PAs. 
\vspace{-1em}
\subsection{Channel Modelling}
In the considered framework, the transmitted signals experience different propagation paths depending on the receiving node. In particular, the signals intended for Alice propagate first through the wireless medium and subsequently through the corresponding waveguide segment toward the BS, whereas Willie directly observes the wireless transmissions over the air. Accordingly, the relevant channels are characterized as follows
\subsubsection{\textbf{Channels associated with Alice}}
The signal transmitted by $U_{B,n}$ (or $U_{C,n}$) first propagates through the wireless medium to reach $U_{P,n}$. After reception at the PA, the superposed signal traverses the corresponding waveguide segment toward the associated feed point at the BS. Hence, the Alice-side propagation can be categorized into the following two channel components
\begin{itemize}
    \item \textbf{Wireless Channel:} The wireless channel characterizes over-the-air propagation from Bob (or Carol) to Alice. For the intra-segment links, the geometry-dependent LoS propagation model is adopted to represent the estimated channel, whereas the inter-segment links are characterized by NLoS fading and remain unknown at Alice. Accordingly, the estimated intra-segment channel between $U_{R,n}$ and $U_{P,n}$, and the inter-segment channel between $U_{R,i}$ and $U_{P,n}$, where $R\in\{B,C\}$ and $i\neq n$, are respectively expressed as
    \begin{align}
        \widetilde{h}_{R,n}^{P,n}&=\frac{\xi e^{-j\frac{2\pi}{\lambda}\left(\left\lVert \Phi_{P,n}-\Phi_{R,n} \right\rVert\right)}}{\left\lVert \Phi_{P,n}-\Phi_{R,n} \right\rVert},\\
        h_{R,i}^{P,n}&=\frac{g_{R,i}^{P,n}}{\sqrt{\left\lVert \Phi_{P,n}-\Phi_{R,i} \right\rVert^{\nu}}},
    \end{align}
    where $\xi=\frac{\lambda}{4\pi}$ denotes the free-space channel gain coefficient at a reference distance of 1m, with $\lambda=\frac{c}{f_C}$. Here, $c$ is the speed of light and $f_C$ denotes the carrier frequency. Moreover, $\nu$ represents the NLoS path-loss exponent, while $g_{R,i}^{P,n} \sim\EuScript{CN}\left\{0,\Omega_{R,i}^{P,n}\right\}$ denotes the small-scale fading coefficient associated with the inter-segment NLoS link where $\Omega_{R,i}^{P,n}$ is the normalised average power of the corresponding NLoS small-scale fading coefficient \cite{wang2026performance}.
    \item \textbf{In-waveguide Channel:} The in-waveguide channel characterizes signal propagation from the PA to the BS through the corresponding waveguide segment, accounting for attenuation and phase variation. The attenuation is governed by the coefficient $\kappa$ (in dB/m), while the phase variation depends on the guided wavelength $\lambda_G=\frac{\lambda}{\eta_{eff}}$, where $\eta_{eff}$ denotes the effective refractive index and $\lambda=\frac{c}{f_C}$ is the free-space wavelength. Since the in-waveguide channel is deterministic, it is assumed to be perfectly known, as commonly considered in PASS-based systems \cite{11495176}. Accordingly, the in-waveguide channel for $\EuScript{S}_n$ is expressed as
\begin{equation}
    h_{\omega,n}=\underbrace{10^{-\frac{\kappa\left\lVert \Phi_{P,n}-\Phi_{F,n} \right\rVert}{20}}}_{\text{Attenuation factor}}\underbrace{e^{-j\frac{2\pi}{\lambda}\left(\eta_{\text{eff}}\left\lVert \Phi_{P,n}-\Phi_{F,n} \right\rVert\right)}}_{\text{Phase Shift}}.
\end{equation}
Here, we have considered that $\eta_{eff}$ and $\kappa$ are the same for each waveguide segment.
\end{itemize}
\subsubsection{\textbf{Channels associated with Willie}} Willie directly receives the over-the-air transmissions originating from both intra-segment and inter-segment users, as indicated by the red arrows and black dashed arrows in \figurename{~\ref{GeneralDiagram}}, respectively. Accordingly, the user-to-Willie wireless channels can be expressed as
\begin{equation}
\label{NLoS_Willie_Channel}
    h_{R,i}^{W,n}=\frac{g_{R,i}^{W,n}}{\sqrt{\left\lVert \Phi_{W,n}-\Phi_{R,i} \right\rVert^{\nu}}},
\end{equation}
where $g_{R,i}^{W,n}\sim\EuScript{CN}\{0,\Omega_{R,i}^{W,n}\}$ denotes the small-scale fading coefficient of the user-to-Willie link where $\Omega_{R,i}^{W,n}$ is the normalised average power of the corresponding NLoS small-scale fading coefficient. Here, $i=n$ corresponds to the intra-segment link between $U_{R,n}$ and $U_{W,n}$, whereas $i\neq n$ represents an inter-segment link from $U_{R,i}$ to $U_{W,n}$. At Willie, pCSI of the intra-segment links is assumed to ensure a conservative secrecy and covertness analysis. This assumption provides Willie with more favourable eavesdropping and detection capabilities, thereby avoiding an overestimation of the achievable secrecy and covert performance.
\subsection{Channel Blockage Modelling and CSI Assumptions}
\subsubsection{\textbf{Channel Blockage Modelling}}
To account for the blockage on the wireless links, a binary random variable $\alpha_{R,i}^{X,n}\in\{0,1\}$, where $R\in\{B,C\}$ and $X\in\{P,W\}$, is introduced. Specifically, $\alpha_{R,i}^{X,n}=1$ indicates that the corresponding wireless link is available, whereas $\alpha_{R,i}^{X,n}=0$ denotes the blocked-link condition. Following the distance-dependent blockage model adopted in EDMA-based systems \cite{ding2025environment,Bhat2612:Performance}, the link-availability probability is expressed as
\begin{equation}
\label{Blockage_Modelling}
    \text{Pr}\left\{\alpha_{R,i}^{X,n}=1\right\}=e^{-\phi_{R,i}^{X,n}\lVert\Phi_{X,n}-\Phi_{R,i}\rVert^2},
\end{equation}
where $\phi_{R,i}^{X,n}$ denotes the blockage parameter associated with the corresponding link, with a larger value indicating a higher likelihood of blockage. Owing to the relatively short distances between the nodes within the same segment, the intra-segment links are assumed to be always available and to contribute to the received signal. Thus, $\alpha_{R,n}^{P,n}=\alpha_{R,n}^{W,n}=1$.
\par Moreover, it follows from \eqref{Blockage_Modelling} that the link-availability probability decreases as the separation between the corresponding nodes increases. Hence, transmissions originating from non-adjacent segments are more likely to be blocked and also experience larger propagation loss. Consequently, the adjacent segments are considered to be the dominant sources of inter-segment interference, while the contribution from farther segments is neglected owing to its comparatively weaker effect. Accordingly, let $\EuScript{A}_n\subseteq\{1,\ldots,N\}$ denote the set of segments adjacent to $\EuScript{S}_n$, defined as
\begin{equation}
\EuScript{A}_n=
    \begin{cases}
    \emptyset, & N=1,\\
    \{2\}, & N>1,\text{ } n=1,\\
    \{n-1,n+1\}, & N>1,\text{ } 1< n< N,\\
    \{N-1\}, & N>1,\text{ } n=N.
    \end{cases}
\end{equation}
\subsubsection{\textbf{CSI Assumptions}} It is assumed that the receiving nodes possess channel knowledge only for the intra-segment links, whereas the instantaneous CSI of the inter-segment links is unavailable and the corresponding signals are treated as interference. Accordingly, $U_{P,n}$ and $U_{W,n}$ possess CSI only for the channels associated with $U_{R,n}$. Furthermore, ipCSI is considered for the intra-segment links at Alice to account for practical channel-estimation errors at the legitimate receiver, whereas pCSI of the corresponding intra-segment links is assumed at Willie in order to adopt a conservative secrecy and covertness analysis. This assumption provides Willie with more favourable detection and eavesdropping capabilities and therefore avoids overstating the achievable secrecy and covert performance.
\par During the channel-estimation phase, Bob and Carol within each segment employ mutually orthogonal pilot sequences selected from a segment-specific pilot set. The pilot sets assigned to the adjacent segments are orthogonal to that of $\EuScript{S}_n$. However, these sets and their corresponding pilot assignments are not available at $\EuScript{S}_n$. Consequently, after correlating the received training signal with the locally known pilots, the pilot components originating from the adjacent segments are suppressed, while dedicated channel estimation of the corresponding inter-segment links cannot be performed. Therefore, $U_{P,n}$ acquires instantaneous CSI only for the intra-segment links, whereas the inter-segment links remain unknown and are treated as interference during data transmission. Accordingly, the equivalent estimated intra-segment channel from $U_{R,n}$ to the BS through $U_{P,n}$ is defined as $\widetilde{H}_{R,n}^{P,n}=h_{\omega,n}\widetilde{h}_{R,n}^{P,n}$.
\par To account for ipCSI of the intra-segment links at Alice, a correlation-based channel-estimation error model is adopted \cite{11005423}. Accordingly, the actual equivalent channel between $U_{R,n}$ and the BS through $U_{P,n}$ is expressed as
\begin{equation}
\label{ipCSI_Equation}
    H_{R,n}^{P,n}=\rho_{R,n} \widetilde{H}_{R,n}^{P,n}+\sqrt{1-\rho_{R,n}^2}e_{R,n}^{P,n},
\end{equation}
where $0<\rho_{R,n}\leq 1$ denotes the channel correlation coefficient characterizing the channel-estimation quality, and $e_{R,n}^{P,n}\sim\EuScript{CN}\left\{0,\mathbb{E}\left[\left\lvert H_{R,n}^{P,n}\right\rvert^2\right]\right\}$ represents an independent complex Gaussian error term, independent of the estimated channel. Similar correlation-based representations between the actual and estimated channels have been employed for ipCSI modelling, while the correlation between the actual and estimated channel powers under pilot-symbol-assisted modulation (PSAM) is characterized in \cite{toan2018performance,809706}. Since the considered intra-segment channels are assumed to remain quasi-static over the pilot and subsequent data-transmission interval, the channel-estimation quality is characterized using the average equivalent channel power. Accordingly, the average pilot SNR for $U_{R,n}$ is defined as $\Gamma_{Pil,R,n}=\frac{\EuScript{P}_{R,n}\mathbb{E}\left[\left\lvert H_{R,n}^{P,n}\right\rvert^2\right]}{\delta_{R,n}\sigma_{P,n}^2},$ where $\EuScript{P}_{R,n}$ is the transmit power of $U_{R,n}$, $\sigma_{P,n}^2$ is the noise variance at $U_{P,n}$ in $\EuScript{S}_n$, and $\delta_{R,n}\geq 1$ denotes the pilot-estimation degradation factor, which represents the loss in channel-estimation quality due to practical imperfections. The case of $\delta_{R,n}= 1$ corresponds to the ideal pilot-estimation condition, whereas larger values indicate poorer estimation quality. Thus, the complex channel correlation coefficient is expressed as $\rho_{R,n}=\sqrt{\frac{\EuScript{N}_{R,n}\EuScript{P}_{R,n}\mathbb{E}\left[\left\lvert H_{R,n}^{P,n}\right\rvert^2\right]}{\EuScript{N}_{R,n}\EuScript{P}_{R,n}\mathbb{E}\left[\left\lvert H_{R,n}^{P,n}\right\rvert^2\right]+\delta_{R,n}\sigma_{P,n}^2}},$ where $\EuScript{N}_{R,n}$ denotes the number of pilot symbols employed for estimating the corresponding intra-segment channel. Accordingly, the effective channel-estimation error variance is given by $\theta_{e,R,n}^2=\left(1-\rho_{R,n}^2\right)\mathbb{E}\left[\left\lvert H_{R,n}^{P,n}\right\rvert^2\right]$.
\subsection{Signal Modelling}
Let $x_{R,n}$ denote the information data generated by $U_{R,n}$ and let $s_{R,n}$ denote the corresponding modulated symbol, normalized such that $\mathbb{E}\left[\lvert s_{R,n} \rvert^2\right]=1$. In each segment, Bob is assumed to transmit continuously, whereas Carol transmits intermittently to preserve communication covertness. This is motivated by the fact that continuous activity of Carol could create a persistent additional signal component at Willie, thereby making the transmission behaviour of $U_{C,n}$ easier to identify. Hence, Carol alternates between active and silent states, and Willie attempts to determine whether a covert transmission is present in a given transmission interval. Accordingly, a binary activity indicator $\beta_{C,n}\in\{0,1\}$ is introduced for $U_{C,n}$, where $\beta_{C,n}=1$ and $\beta_{C,n}=0$ represent the active and silent states, respectively. Now, let $p_{act,n}$ denote the covert user activity probability of $\EuScript{S}_n$, such that
\begin{equation}
\label{Covert_Activity}
    \text{Pr}\{\beta_{C,n}=1\}=p_{act,n}, \quad \text{Pr}\{\beta_{C,n}=0\}=1-p_{act,n}.
\end{equation}
It should be noted that the use of an activity probability is also consistent with \cite{10459011}, where a prior transmission probability is introduced for the covert user. For simplicity, we consider $p_{act}=p_{act,n}=p_{act,n-1}=p_{act,n+1}$. The activity probability $p_{act}$ is assumed to be known at Willie of $\EuScript{S}_n$ and therefore represents the prior probability of $\EuScript{H}_1$, while $1-p_{act}$ represents the prior probability of $\EuScript{H}_0$. Furthermore, owing to the different security requirements of Bob and Carol, Bob transmits with a relatively higher power to support reliable secrecy transmission, whereas Carol employs a comparatively lower transmit power to reduce the likelihood of detection at Willie and thus, ensure covertness \cite{10459011,10938035}. Accordingly, the transmit powers are selected such that $\EuScript{P}_{B,n}>\EuScript{P}_{C,n}$. Now, based on the transmission activity of $U_{C,n}$, two transmission hypotheses are considered
\begin{equation}
\begin{split}
    &\EuScript{H}_0: \beta_{C,n}=0 \text{ (i.e. Carol does not transmit)},\\
    &\EuScript{H}_1: \beta_{C,n}=1 \text{ (i.e. Carol transmits)}.
\end{split}
\end{equation}
The received signal models, along with the corresponding decoding procedures and SINRs at Alice and Willie, are presented next.
\subsubsection{\textbf{Received Signal at Alice of $\EuScript{S}_n$}}
The received signal at $U_{P,n}$ corresponding to $\EuScript{S}_n$ can be expressed as
\begin{equation}
\!\!\!\!\EuScript{Y}_{P,n}\!=\!
    \begin{cases}\!\!
        \sqrt{\EuScript{P}_{B,n}}H_{B,n}^{P,n}s_{B,n}+\mathcal{I}_{P,n}+z_{P,n}, & \EuScript{H}_0,\\
        \sqrt{\EuScript{P}_{B,n}}H_{B,n}^{P,n}s_{B,n}\!+\!\sqrt{\EuScript{P}_{C,n}}H_{C,n}^{P,n}s_{C,n}\!+\!\mathcal{I}_{P,n}+z_{P,n}, & \EuScript{H}_1,
    \end{cases}
\end{equation}
where $z_{P,n}\sim\EuScript{CN}\{0,\sigma_{P,n}^2\}$ denotes the additive white Gaussian noise (AWGN) at $U_{P,n}$, while $\mathcal{I}_{P,n}$ represents the aggregate inter-segment interference contributed by the adjacent segments. Since $\EuScript{A}_n\subseteq\{1,\ldots,N\}$ contains the indices of the segments adjacent to $\EuScript{S}_n$, $\mathcal{I}_{P,n}$ is given by
\begin{equation}
    \mathcal{I}_{P,n}=h_{\omega,n}\sum_{i\in\EuScript{A}_n}\left(\alpha_{B,i}^{P,n}h_{B,i}^{P,n}\sqrt{\EuScript{P}_{B,i}}s_{B,i}+\beta_{C,i}\alpha_{C,i}^{P,n}h_{C,i}^{P,n}\sqrt{\EuScript{P}_{C,i}}s_{C,i}\right).
\end{equation}
Since $\EuScript{H}_0$ and $\EuScript{H}_1$ characterize only the transmission state of $U_{C,n}$, the aggregate inter-segment interference $\mathcal{I}_{P,n}$ is present under both hypotheses. Now, under $\EuScript{H}_0$, only Bob transmits within $\EuScript{S}_n$. Thus, the SINR for decoding the secrecy signal of $U_{B,n}$ at $U_{P,n}$ is given by
\begin{equation}
\label{Gamma_BnP0}
    \Gamma_{B,n}^{P,0}=\frac{\EuScript{P}_{B,n}\rho_{B,n}^2\left\lvert \widetilde{H}_{B,n}^{P,n}\right\rvert^2}{\EuScript{P}_{B,n}\theta_{e,B,n}^2+\EuScript{I}_{P,n}+\sigma_{P,n}^2},
\end{equation}
where $\EuScript{I}_{P,n}$ denotes the aggregate inter-segment interference power and can be represented as
\begin{equation}
\label{I_Pn}
    \EuScript{I}_{P,n}=\left\lvert h_{\omega,n}\right\rvert^2\sum_{i\in\EuScript{A}_n}\left(\alpha_{B,i}^{P,n}\left\lvert h_{B,i}^{P,n}\right\rvert^2\EuScript{P}_{B,i}+\beta_{C,i}\alpha_{C,i}^{P,n}\left\lvert h_{C,i}^{P,n}\right\rvert^2\EuScript{P}_{C,i}\right).
\end{equation}
Under $\EuScript{H}_1$, Carol also transmits and Bob’s higher-power secrecy signal is decoded first while treating Carol’s covert signal as interference. Therefore, the corresponding instantaneous SINR is given as
\begin{equation}
\label{Gamma_BnP1}
    \!\!\!\Gamma_{B,n}^{P,1}\!=\!\frac{\EuScript{P}_{B,n}\rho_{B,n}^2\left\lvert \widetilde{H}_{B,n}^{P,n}\right\rvert^2}{\EuScript{P}_{C,n}\rho_{C,n}^2\left\lvert \widetilde{H}_{C,n}^{P,n}\right\rvert^2\!+\!\EuScript{P}_{B,n}\theta_{e,B,n}^2\!+\!\EuScript{P}_{C,n}\theta_{e,C,n}^2\!+\!\EuScript{I}_{P,n}+\sigma_{P,n}^2}.
\end{equation}
Following the decoding of Bob's secrecy signal, SIC is performed at Alice, and the instantaneous SINR for decoding Carol's covert signal is given by
\begin{equation}
\label{Gamma_CnP1}
    \hspace{-1em}\Gamma_{C,n}^{P,1}\!=\!\frac{\EuScript{P}_{C,n}\rho_{C,n}^2\left\lvert \widetilde{H}_{C,n}^{P,n}\right\rvert^2}{\Delta_{P,n}\rho_{B,n}^2\EuScript{P}_{B,n}\left\lvert \widetilde{H}_{B,n}^{P,n}\right\rvert^2\!\!+\!\EuScript{P}_{B,n}\theta_{e,B,n}^2\!+\!\EuScript{P}_{C,n}\theta_{e,C,n}^2\!+\!\EuScript{I}_{P,n}\!+\!\sigma_{P,n}^2}\!,
\end{equation}
where $\Delta_{P,n}\in[0,1]$ denotes the residual interference coefficient caused by imperfect cancellation of the secrecy signal component.
\vspace{-0.5em}
\subsubsection{\textbf{Received Signal at Willie of $\EuScript{S}_n$}}
Since Willie directly observes the over-the-air transmissions, the received signal at $U_{W,n}$ under the two hypotheses can be expressed as
\begin{equation}
\EuScript{Y}_{W,n}=
    \begin{cases}
        \sqrt{\EuScript{P}_{B,n}}h_{B,n}^{W,n}s_{B,n}+\mathcal{I}_{W,n}+z_{W,n}, & \!\!\EuScript{H}_0,\\
        \sqrt{\EuScript{P}_{B,n}}h_{B,n}^{W,n}s_{B,n}+\sqrt{\EuScript{P}_{C,n}}h_{C,n}^{W,n}s_{C,n}+\mathcal{I}_{W,n}+z_{W,n}, & \!\!\EuScript{H}_1,
    \end{cases}
\end{equation}
where $z_{W,n}\sim\EuScript{CN}\{0,\sigma_{W,n}^2\}$ denotes the AWGN  at $U_{W,n}$, while $\mathcal{I}_{W,n}$ denotes the aggregate inter-segment interference contributed by the adjacent segments and is given by
\begin{equation}
    \label{Interference_Wn}
    \mathcal{I}_{W,n}=\sum_{i\in\EuScript{A}_n}\left(\alpha_{B,i}^{W,n}h_{B,i}^{W,n}\sqrt{\EuScript{P}_{B,i}}s_{B,i}+\beta_{C,i}\alpha_{C,i}^{W,n}h_{C,i}^{W,n}\sqrt{\EuScript{P}_{C,i}}s_{C,i}\right).
\end{equation}
Accordingly, the corresponding aggregate inter-segment interference power, averaged over the transmitted symbols, can be represented as
\begin{equation}
    \label{Interference_Power_Wn}\EuScript{I}_{W,n}=\sum_{i\in\EuScript{A}_n}\left(\alpha_{B,i}^{W,n}\left\lvert h_{B,i}^{W,n}\right\rvert^2\EuScript{P}_{B,i}+\beta_{C,i}\alpha_{C,i}^{W,n}\left\lvert h_{C,i}^{W,n}\right\rvert^2\EuScript{P}_{C,i}\right).
\end{equation}
Since Bob continuously transmits with a higher transmit power, Willie first decodes Bob's secrecy signal before processing Carol's covert transmission. Under $\EuScript{H}_0$, the instantaneous SINR for decoding Bob's signal at Willie is given by
\begin{equation}
\label{Gamma_BnW0}
    \Gamma_{B,n}^{W,0}=\frac{\EuScript{P}_{B,n}\left\lvert h_{B,n}^{W,n}\right\rvert^2}{\EuScript{I}_{W,n}+\sigma_{W,n}^2}.
\end{equation}
Under $\EuScript{H}_1$, Carol's transmission additionally acts as interference while Willie decodes Bob. Hence, instantaneous SINR for decoding Bob's signal at Willie can be given as
\begin{equation}
\label{Gamma_BnW1}
    \Gamma_{B,n}^{W,1}=\frac{\EuScript{P}_{B,n}\left\lvert h_{B,n}^{W,n}\right\rvert^2}{\EuScript{P}_{C,n}\left\lvert h_{C,n}^{W,n}\right\rvert^2+\EuScript{I}_{W,n}+\sigma_{W,n}^2}.
\end{equation}
Following the decoding of Bob's secrecy signal, Willie performs SIC and utilizes the residual observation to determine whether Carol is transmitting. To adopt a conservative assumption for covert communication, Willie is assumed to successfully decode Bob’s secrecy signal and perfectly cancel it before detecting Carol’s transmission. Accordingly, the post-SIC observation is given as
\begin{equation}
\label{After_SIC_Willie}
\widehat{\EuScript{Y}}_{W,n}=
    \begin{cases}
        \mathcal{I}_{W,n}+z_{W,n}, & \EuScript{H}_0,\\
        \sqrt{\EuScript{P}_{C,n}}h_{C,n}^{W,n}s_{C,n}+\mathcal{I}_{W,n}+z_{W,n}, & \EuScript{H}_1.
    \end{cases}
\end{equation}
It should be noted that, unlike \cite{10459011}, even after perfect SIC of Bob's signal, the inter-segment interference remains present in the residual observation and therefore contributes to the uncertainty in distinguishing between $\EuScript{H}_0$ and $\EuScript{H}_1$. Under $\EuScript{H}_1$, the instantaneous SINR associated with Carol's covert signal at Willie after SIC is therefore given by
\begin{equation}
    \Gamma_{C,n}^{W,1}=\frac{\EuScript{P}_{C,n}\left\lvert h_{C,n}^{W,n}\right\rvert^2}{\EuScript{I}_{W,n}+\sigma_{W,n}^2}.
\end{equation}
The post-SIC observations under $\EuScript{H}_0$ and $\EuScript{H}_1$ are subsequently utilized for Willie’s binary hypothesis testing.
\subsubsection{\textbf{Instantaneous Secrecy and Covert Rates}}
Based on the instantaneous SINRs of Bob at Alice and Willie, the instantaneous secrecy rate of $U_{B,n}$ under hypothesis $\EuScript{H}_q, q\in\{0,1\}$, is given by
\begin{equation}
\label{Instant_Secrecy_Rate}
    C_{sec,n}^{(q)}=\left[\log_2\left(1+\Gamma_{B,n}^{P,q}\right)-\log_2\left(1+\Gamma_{B,n}^{W,q}\right)\right]^{+},
\end{equation}
where $[x]^{+}=\max(x,0)$. On the other hand, since Carol transmits intermittently, the instantaneous covert rate can be given as
\begin{equation}
    \label{Instant_Covert_Rate}
    C_{cov,n}=
    \begin{cases}
        0, & \EuScript{H}_0,\\
        \log_2\left(1+\Gamma_{C,n}^{P,1}\right) & \EuScript{H}_1.\\
    \end{cases}
\end{equation}
It should be noted that the covert transmission rate characterizes the achievable information rate of Carol’s legitimate link, whereas the corresponding level of covertness is independently characterized by the minimum DEP at Willie.
\section{Performance Analysis}
\label{PerformanceAnalysis}
This section presents the performance analysis of the considered framework in terms of SOP, DEP, ASC, and CR. The performance metrics depend jointly on random user locations, distance-dependent blockage states, inter-segment NLoS fading, and adjacent-segment covert-user activity, resulting in numerous possible network configurations. Accounting analytically for all such configurations would lead to highly complicated expressions. Therefore, a semi-analytical approach is adopted, wherein the random network configuration, including the resulting inter-segment interference, is numerically realized, while the remaining intra-segment wireless fading is analytically characterized. The resulting conditional performance expressions are subsequently averaged over independent network realizations to obtain the overall performance metrics. Furthermore, before proceeding with the individual performance metrics, we first characterize the distribution of $\left\lvert\widetilde{H}_{R,n}^{P,n}\right\rvert^2$, which will be utilized in the subsequent analysis and is presented in Lemma \ref{Lemma_1}. Since the derived channel-power distributions are piecewise defined over finite intervals, composite midpoint quadrature is adopted to evaluate the resulting integrals using the midpoints of uniformly partitioned integration intervals, providing a simple and consistent numerical framework for the semi-analytical performance analysis.
\begin{lemma}
\label{Lemma_1}
    For a node in $\EuScript{S}_n$, CDF and PDF of $\left\lvert\widetilde{H}_{R,n}^{P,n}\right\rvert^2$ are respectively given in \eqref{CDF_Channel} and \eqref{PDF_Channel} (next page top), where $\EuScript{E}(x)=\frac{\xi^2\left\lvert h_{\omega,n}\right\rvert^2}{x}-d^2, L_{min}=\min(L_S,W),$ and $L_{max}=\max(L_S,W)$. Furthermore, $\EuScript{Q}_1(x)=L_S\sqrt{\EuScript{E}(x)-\frac{L_S^2}{4}}+W\sqrt{\EuScript{E}(x)-\frac{W^2}{4}}+2\EuScript{E}(x)\left[\sin^{-1}\left(\frac{L_S}{2\sqrt{\EuScript{E}(x)}}\right)+\sin^{-1}\left(\frac{W}{2\sqrt{\EuScript{E}(x)}}\right)-\frac{\pi}{2}\right]$, and $\EuScript{Q}_2(x)=L_{min}\sqrt{\EuScript{E}(x)-\frac{L_{min}^2}{4}}+2\EuScript{E}(x)\sin^{-1}\left(\frac{L_{min}}{2\sqrt{\EuScript{E}(x)}}\right)$.
    \begin{equation}
    \label{CDF_Channel}
        \EuScript{F}_{\left\lvert\widetilde{H}_{R,n}^{P,n}\right\rvert^2}(x)=
        \begin{cases}
            0, & \EuScript{E}(x)\geq \frac{L_S^2+W^2}{4},\\
            1-\frac{\EuScript{Q}_1(x)}{L_SW}, & \frac{L_{max}^2}{4}<\EuScript{E}(x)<\frac{L_S^2+W^2}{4},\\
            1-\frac{\EuScript{Q}_2(x)}{L_SW} & \frac{L_{min}^2}{4}<\EuScript{E}(x)<\frac{L_{max}^2}{4},\\
            1-\frac{\pi\EuScript{E}(x)}{L_SW}, & 0<\EuScript{E}(x)\leq \frac{L_{min}^2}{4},\\
            1, & \EuScript{E}(x)\leq 0.\\
        \end{cases}
    \end{equation}
\begin{figure*}[ht]
    \begin{equation}
    \label{PDF_Channel}
        f_{\left\lvert\widetilde{H}_{R,n}^{P,n}\right\rvert^2}(x)=\frac{\xi^2\left\lvert h_{\omega,n}\right\rvert^2}{L_SWx^2}
        \begin{cases}
            \pi, & 0<\EuScript{E}(x)\leq \frac{L_{min}^2}{4},\\
            2\sin^{-1}\left(\frac{L_{min}}{2\sqrt{\EuScript{E}(x)}}\right), & \frac{L_{min}^2}{4}<\EuScript{E}(x)\leq \frac{L_{max}^2}{4},\\
            2\left[\sin^{-1}\left(\frac{L_S}{2\sqrt{\EuScript{E}(x)}}\right)+\sin^{-1}\left(\frac{W}{2\sqrt{\EuScript{E}(x)}}\right)-\frac{\pi}{2}\right], & \frac{L_{max}^2}{4}<\EuScript{E}(x)<\frac{L_S^2+W^2}{4},\\
            0, & \text{otherwise.}
        \end{cases}
    \end{equation}
    \hrule
    \end{figure*}

\begin{proof}
    We have $\left\lvert\widetilde{H}_{R,n}^{P,n}\right\rvert^2=\frac{\xi^2\left\lvert h_{\omega,n}\right\rvert^2}{\left\lVert \Phi_{P,n}-\Phi_{R,n} \right\rVert^2}$. Accordingly, its CDF can be given as
    \begin{equation}
    \begin{split}
        \EuScript{F}_{\left\lvert\widetilde{H}_{R,n}^{P,n}\right\rvert^2}(x)&=\Pr\left\{\left\lvert\widetilde{H}_{R,n}^{P,n}\right\rvert^2\!\leq\! x\right\}\\&=\Pr\left\{\left\lVert \Phi_{P,n}-\Phi_{R,n} \right\rVert^2> \frac{\xi^2\left\lvert h_{\omega,n}\right\rvert^2}{x}\right\}\\&=\Pr\left\{(x_{R,n}-x_{P,n}^2)+y_{R,n}^2> \EuScript{E}(x)\right\}.
    \end{split}
    \end{equation}
    Since $U_{R,n}$ is uniformly distributed within $\EuScript{S}_n$ and the corresponding PA is positioned at the center of the segment, the above probability is determined by the portion of the rectangular region lying outside a circle of radius $\sqrt{\EuScript{E}(x)}$. Depending on the circle radius relative to the segment dimensions, three geometric cases arise: the circle lies entirely within the rectangular region, intersects its shorter sides, or intersects both pairs of sides. Exploiting the symmetry of the rectangular region and evaluating the corresponding intersection areas yields the CDF in \eqref{CDF_Channel}. Subsequently, differentiating \eqref{CDF_Channel} with respect to (wrt) $x$ gives the PDF in \eqref{PDF_Channel}.
\end{proof}
\end{lemma}
\subsection{Secrecy Outage Analysis}
For a target secrecy rate $\EuScript{R}_{sec,Th}(>0)$, the secrecy outage probability at $\EuScript{S}_n$ can be expressed as $\mathbb{P}_{sec,n}=(1-p_{act})\mathbb{P}_{sec,n}^{(0)}+p_{act}\mathbb{P}_{sec,n}^{(1)},$ where $p_{act}$ is the covert user activity probability as discussed in \eqref{Covert_Activity} and $\mathbb{P}_{sec,n}^{(q)}, q\in\{0,1\}$. Using the instantaneous secrecy rate in \eqref{Instant_Secrecy_Rate}, $\mathbb{P}_{sec,n}^{(q)}$ can be expressed as
\vspace{-0.25em}
\begin{equation}
\label{SOP}
\begin{split}
    \mathbb{P}_{sec,n}^{(q)}&=\Pr\left\{C_{sec,n}^{(q)}<\EuScript{R}_{sec,Th}\right\}\\&=\Pr\left\{\log_2\left(\frac{1+\Gamma_{B,n}^{P,q}}{1+\Gamma_{B,n}^{W,q}}\right)<\EuScript{R}_{sec,Th}\right\}\\
    &=\Pr\left\{\Gamma_{B,n}^{P,q}<\Theta_{sec}\left(1+\Gamma_{B,n}^{W,q}\right)-1\right\},
\end{split}
\end{equation}
where $\Theta_{sec}=2^{\EuScript{R}_{sec,Th}}$. The evaluation of \eqref{SOP} involves the estimated Alice-side channel powers, intra-segment fading at Willie, and the aggregate inter-segment interference, all of which depend on the random network configuration. In particular, the estimated channel powers and the corresponding Willie-side channel statistics share the same user locations, resulting in statistical dependence. To account for this dependence, we employ a semi-analytical approach in which the estimated channel powers are characterized using the piecewise PDFs derived in Lemma 1, while the remaining conditional averages are evaluated using the corresponding network configurations. In the following, we discuss how $\mathbb{P}_{sec,n}^{(0)}$ and $\mathbb{P}_{sec,n}^{(1)}$ can be obtained, respectively.
\subsubsection{\textbf{Under Hypothesis $\EuScript{H}_0 \text{ (i.e. $q=0$)}$}} Under hypothesis $\EuScript{H}_0$, \eqref{Instant_Secrecy_Rate} can be given as
\begin{equation}
    C_{sec,n}^{(0)}=\left[\log_2\left(1+\Gamma_{B,n}^{P,0}\right)-\log_2\left(1+\Gamma_{B,n}^{W,0}\right)\right]^{+}.
\end{equation}
Accordingly, using \eqref{Gamma_BnP0} and \eqref{SOP}, the SOP under $\EuScript{H}_0$ is given as
\begin{equation}
\begin{split}
\label{30}
    \mathbb{P}_{sec,n}^{(0)}&=\Pr\left\{\Gamma_{B,n}^{P,0}<\Theta_{sec}\left(1+\Gamma_{B,n}^{W,0}\right)-1\right\}\\&=\Pr\left\{\left\lvert \widetilde{H}_{B,n}^{P,n}\right\rvert^2<\tau_{B,n}^{(0)}(x)\right\},
\end{split}
\end{equation}
where $\tau_{B,n}^{(0)}(x)=\frac{\EuScript{P}_{B,n}\theta_{e,B,n}^2+\EuScript{I}_{P,n}+\sigma_{P,n}^2}{\EuScript{P}_{B,n}\rho_{B,n}^2}\left[\Theta_{sec}\left(1+\frac{\EuScript{P}_{B,n}x}{\EuScript{I}_{W,n}+\sigma_{W,n}^2}\right)-1\right]$ with $x=\left\lvert h_{B,n}^{W,n}\right\rvert^2$. Thus, applying the law of total probability to \eqref{30}, we obtain
\begin{equation}
\label{31B}
    \mathbb{P}_{sec,n}^{(0)}=\int_{H_{min}}^{H_{max}} f_{\left\lvert\widetilde{H}_{B,n}^{P,n}\right\rvert^2}(u)\times \Pr\left\{u<\tau_{B,n}^{(0)}(x)\Bigg|\left\lvert\widetilde{H}_{B,n}^{P,n}\right\rvert^2=u\right\}du,
\end{equation}
where $H_{min}=\frac{\xi^2\left\lvert h_{\omega,n}\right\rvert^2}{d^2+\frac{L_S^2+W^2}{4}}$ and $H_{max}=\frac{\xi^2\left\lvert h_{\omega,n}\right\rvert^2}{d^2}$ denote the lower and upper bounds of the channel-power. Now, for a fixed network geometry, $x$ follows an exponential distribution with mean $\mu_{B,n}^{W,n}=\frac{\Omega_{B,n}^{W,n}}{\left\lVert \Phi_{W,n}-\Phi_{B,n} \right\rVert^{\nu}}$. Hence, averaging the conditional outage event over Bob-to-Willie fading yields
\begin{equation}
\label{32}
    \mathbb{P}_{sec,n}^{(0)}|\EuScript{G},\Phi_{B,n}\!=\!e^{-\frac{\EuScript{I}_{W,n}+\sigma_{W,n}^2}{\EuScript{P}_{B,n}\mu_{B,n}^{W,n}}\left[\frac{1}{\Theta_{sec}}\left(1+\frac{\EuScript{P}_{B,n}\rho_{B,n}^2\left\lvert \widetilde{H}_{B,n}^{P,n}\right\rvert^2}{\EuScript{P}_{B,n}\theta_{e,B,n}^2+\EuScript{I}_{P,n}+\sigma_{P,n}^2}\right)-1\right]^+}\!,
\end{equation}
where $\EuScript{G}$ represents the network configuration comprising the user positions in the adjacent segments, blockage states, CSI parameters, position of Willie in the considered segment, and inter-segment interference. The position of Bob and Carol in the considered segment remain random and are averaged over using the channel-power distributions derived in Lemma \ref{Lemma_1}. The Willie-channel mean in \eqref{32} is evaluated using the corresponding Bob and Willie positions, while the CSI parameters are determined from the average equivalent channel power. Now, substituting \eqref{32} in \eqref{31B}, we obtain
\begin{equation}
\label{33}
    \mathbb{P}_{sec,n}^{(0)}=\int_{H_{min}}^{H_{max}} f_{\left\lvert\widetilde{H}_{B,n}^{P,n}\right\rvert^2}(u)\times \mathbb{E}_{\left\lvert\widetilde{H}_{B,n}^{P,n}\right\rvert^2=u}\left[\mathbb{P}_{sec,n}^{(0)}|\EuScript{G},\Phi_{B,n}\right] du.
\end{equation}
The conditional expectation in the above equation is numerically evaluated over the remaining network randomness for a specified estimated Bob-to-Alice channel power, thereby preserving the spatial dependence between the legitimate and Willie-side channel statistics. The above equations are evaluated using the composite midpoint quadrature rule.
\subsubsection{\textbf{Under Hypothesis $\EuScript{H}_1 \text{ (i.e. $q=1$)}$}} Under $\EuScript{H}_1$, Carol also transmits within $\EuScript{S}_n$. Hence, using \eqref{Gamma_BnP1}, \eqref{Gamma_BnW1}, and \eqref{SOP}, the SOP under $\EuScript{H}_1$ can be given as \vspace{-0.5em}
\begin{equation}
\label{34}
    \mathbb{P}_{sec,n}^{(1)}=\Pr\left\{\left\lvert \widetilde{H}_{B,n}^{P,n}\right\rvert^2<\tau_{B,n}^{(1)}\left(x,y,z\right)\right\},
\end{equation}
\begin{figure*}
\begin{equation}
\label{tau_Bn_H1}
    \tau_{B,n}^{(1)}\left(x,y,z\right)=\frac{\left(\EuScript{P}_{C,n}\rho_{C,n}^2z+\EuScript{P}_{B,n}\theta_{e,B,n}^2+\EuScript{P}_{C,n}\theta_{e,C,n}^2+\EuScript{I}_{P,n}+\sigma_{P,n}^2\right)}{\EuScript{P}_{B,n}\rho_{B,n}^2}\left[\Theta_{sec}\left(1+\frac{\EuScript{P}_{B,n}x}{\EuScript{P}_{C,n}y+\EuScript{I}_{W,n}+\sigma_{W,n}^2}\right)-1\right]
\end{equation}
\vspace{-0.5em}
\hrule
\end{figure*}
where $\tau_{B,n}^{(1)}\left(x,y,z\right)$ is given in \eqref{tau_Bn_H1} with $x=\left\lvert h_{B,n}^{W,n}\right\rvert^2$, $y=\left\lvert h_{C,n}^{W,n}\right\rvert^2$, and $z=\left\lvert \widetilde{H}_{C,n}^{P,n}\right\rvert^2$. From Lemma \ref{Lemma_1}, $\left\lvert \widetilde{H}_{B,n}^{P,n}\right\rvert^2$ and $\left\lvert \widetilde{H}_{C,n}^{P,n}\right\rvert^2$ follow the piecewise PDFs given in \eqref{PDF_Channel}. Furthermore, since Bob and Carol are independently distributed within $\EuScript{S}_n$ their estimated channel powers are also independent. Hence, applying the law of total probability to \eqref{34}, the SOP can be written as
\begin{equation}
\label{36}
\begin{split}
    \mathbb{P}_{sec,n}^{(1)}&=\int_{H_{min}}^{H_{max}} \int_{H_{min}}^{H_{max}}f_{\left\lvert\widetilde{H}_{B,n}^{P,n}\right\rvert^2}(u)f_{\left\lvert\widetilde{H}_{C,n}^{P,n}\right\rvert^2}(z)\times\\& \hspace{7mm}\Pr\left\{u<\tau_{B,n}^{(1)}(x,y,z)\Bigg|\left\lvert\widetilde{H}_{B,n}^{P,n}\right\rvert^2=u,\left\lvert\widetilde{H}_{C,n}^{P,n}\right\rvert^2=z\right\}dzdu.
\end{split}
\end{equation}
Now, for a given network configuration $\EuScript{G}$, $\Phi_{B,n}$, and $\Phi_{C,n}$, the Bob-to-Willie and Carol-to-Willie fading powers are independent and exponentially distributed, with respective means $\mu_{B,n}^{W,n}=\frac{\Omega_{B,n}^{W,n}}{\left\lVert \Phi_{W,n}-\Phi_{B,n} \right\rVert^{\nu}}$ and $\mu_{C,n}^{W,n}=\frac{\Omega_{C,n}^{W,n}}{\left\lVert \Phi_{W,n}-\Phi_{C,n} \right\rVert^{\nu}}$. Accordingly, averaging the conditional outage probability over the corresponding exponential fading distributions yields
\begin{equation}
    \mathbb{P}_{sec,n}^{(1)}|\EuScript{G},\Phi_{B,n},\Phi_{C,n}=\frac{e^{-\frac{\EuScript{I}_{W,n}+\sigma_{W,n}^2}{\EuScript{P}_{B,n}\mu_{B,n}^{W,n}}\left[\frac{1+\Gamma_{B,n}^{P,1}(u,z)}{\Theta_{sec}}-1\right]^+}}{1+\frac{\EuScript{P}_{C,n}\mu_{C,n}^{W,n}}{\EuScript{P}_{B,n}\mu_{B,n}^{W,n}}\left[\frac{1+\Gamma_{B,n}^{P,1}(u,z)}{\Theta_{sec}}-1\right]^+}.
\end{equation}
Here, $\Gamma_{B,n}^{P,1}(u,z)$ denotes the SINR in \eqref{Gamma_BnP1}, with the estimated channel powers of Bob and Carol evaluated at $u$ and $z$, respectively. Finally, substituting this in \eqref{36}, we obtain
\begin{equation}
\label{38}
\hspace{-1.75em}\begin{split}
    \mathbb{P}_{sec,n}^{(1)}&\!=\!\int_{H_{min}}^{H_{max}} \int_{H_{min}}^{H_{max}}f_{\left\lvert\widetilde{H}_{B,n}^{P,n}\right\rvert^2}(u)f_{\left\lvert\widetilde{H}_{C,n}^{P,n}\right\rvert^2}(z)\times\\&\hspace{8mm}\mathbb{E}\left[\mathbb{P}_{sec,n}^{(1)}|\EuScript{G}\Bigg|\left\lvert \widetilde{H}_{B,n}^{P,n}\right\rvert^2=u,\left\lvert \widetilde{H}_{C,n}^{P,n}\right\rvert^2=z\right]dzdu.
    \end{split}
\end{equation}
Finally, substituting \eqref{33} and \eqref{38} into the overall SOP expression yields the secrecy outage probability of the considered system. The above equations are evaluated using the composite midpoint quadrature rule.
\subsection{Detection Error Probability}
DEP is characterised by Willie's uncertainty in determining whether Carol of $\EuScript{S}_n$ is transmitting covert information. Since, $\EuScript{P}_{B,n}>\EuScript{P}_{C,n}$, Willie first decodes Bob's higher-power secrecy signal and, following the conservative assumption in \eqref{After_SIC_Willie}, perfectly cancels it before detecting Carol's transmission. Moreover, pCSI of the corresponding intra-segment channels is assumed at Willie. Hence, based on the post-SIC observation in \eqref{After_SIC_Willie}, $U_{W,n}$ performs a binary hypothesis test to distinguish between $\EuScript{H}_0$ and $\EuScript{H}_1$. Considering the prior probabilities of $\EuScript{H}_0$ and $\EuScript{H}_1$, Willie of $\EuScript{S}_n$ employs a radiometer and performs a minimum error Bayesian binary hypothesis test by comparing the received signal energy with a predefined detection threshold $\zeta_n$ which can be given as
\vspace{-0.0em}
\begin{equation}
\EuScript{T}_{W,n}=\frac{1}{K}\sum_{k=1}^{K}\left|\widehat{\EuScript{Y}}_{W,n}(k)\right|^2\underset{\EuScript{D}_0}{\overset{\EuScript{D}_1}{\gtrless}}\zeta_n,
\end{equation}
where $K$ denotes the number of observed symbols, while $\EuScript{D}_0$ and $\EuScript{D}_1$ denote Willie's decisions in favour of $\EuScript{H}_0$ and $\EuScript{H}_1$, respectively. For sufficiently large values of $K$ i.e. $K\rightarrow\infty$, the radiometer output converges to the corresponding average received power. Thus, using \eqref{Interference_Power_Wn} and \eqref{After_SIC_Willie}, $\EuScript{T}_{W,n}$ is expressed as \vspace{-0.0em}
\begin{equation}
\label{T_Wn}
\EuScript{T}_{W,n}=
\begin{cases}
    \EuScript{I}_{W,n}+\sigma_{W,n}^2, & \EuScript{H}_0,\\
    \EuScript{P}_{C,n}\left\lvert h_{C,n}^{W,n}\right\rvert^2+\EuScript{I}_{W,n}+\sigma_{W,n}^2, & \EuScript{H}_1.
\end{cases}
\end{equation}
Accordingly, the DEP at $U_{W,n}$ is given by
\begin{equation}
\label{DEP}
\begin{split}
    \mathbb{P}_{DEP,n}&=(1-p_{act})\Pr\left\{\EuScript{D}_1|\EuScript{H}_0\right\}+p_{act}\Pr\left\{\EuScript{D}_0|\EuScript{H}_1\right\}\\&=(1-p_{act})\mathbb{P}_{FA,n}+p_{act}\mathbb{P}_{MD,n},
    \end{split}
\end{equation}
where $\mathbb{P}_{FA,n}=\Pr\left\{\EuScript{D}_1|\EuScript{H}_0\right\}$ denotes the false alarm probability (FAP) and $\mathbb{P}_{MD,n}=\Pr\left\{\EuScript{D}_0|\EuScript{H}_1\right\}$ denotes the missed detection probability (MDP). In the following, we derive the FAP and MDP, respectively and we consider $\Lambda_n=\zeta_n-\sigma_{W,n}^2$ during the analysis.
\subsubsection{\textbf{False Alarm Probability}}
False alarm occurs when under $\EuScript{H}_0$, Willie decides $\EuScript{D}_1$. Thus, FAP for $\EuScript{S}_n$ is given as
\begin{equation}
\begin{split}    
    \mathbb{P}_{FA,n}&=\Pr\left\{\EuScript{I}_{W,n}+\sigma_{W,n}^2>\zeta_n\right\}\\&=\Pr\left\{\EuScript{I}_{W,n}>\Lambda_n\right\}=1-\EuScript{F}_{\EuScript{I}_{W,n}}(\Lambda_n),
\end{split}
\end{equation}
for $\Lambda_n\geq 0$ and where $\EuScript{F}_{\EuScript{I}_{W,n}}(.)$ denotes the CDF of the aggregate inter-segment interference power at Willie. For $\Lambda_n<0$, $\mathbb{P}_{FA,n}=1$. Now, as discussed in Section \ref{PerformanceAnalysis}, $\EuScript{I}_{W,n}$ jointly depends on the random user locations, inter-segment NLoS fading, blockage states, and Carol activity in the adjacent segments. Hence, obtaining a compact closed-form expression for $\EuScript{F}_{\EuScript{I}_{W,n}}(.)$ is analytically intractable. Therefore, following the adopted semi-analytical framework, its distribution is numerically evaluated over independent realizations of the random network configuration $\EuScript{G}$. Accordingly, let $\widetilde{\EuScript{F}}_{\EuScript{I}_{W,n}}(.)$ denote its CDF obtained over independent realizations of the random network configuration $\EuScript{G}$. Thus, the semi-analytical FAP can be obtained as\vspace{-0.25em}
\begin{equation}
\label{FAP_Final}
    \mathbb{P}_{FA,n}\approx1-\widetilde{\EuScript{F}}_{\EuScript{I}_{W,n}}(\Lambda_n).
\end{equation} 

\vspace{-0.25em}
\subsubsection{\textbf{Missed Detection Probability}}
A missed-detection event occurs when, under $\EuScript{H}_1$, Willie decides $\EuScript{D}_0$. Accordingly, using \eqref{T_Wn}, MDP can be expressed as
\begin{equation}
\label{MDP}
\begin{split}
    \mathbb{P}_{MD,n}&=\Pr\left\{\EuScript{P}_{C,n}\left\lvert h_{C,n}^{W,n}\right\rvert^2+\EuScript{I}_{W,n}+\sigma_{W,n}^2\leq\zeta_n\right\}\\
    &=\Pr\left\{\EuScript{P}_{C,n}\left\lvert h_{C,n}^{W,n}\right\rvert^2+\EuScript{I}_{W,n}\leq\Lambda_n\right\}.
\end{split}
\end{equation}
Now, for a given network configuration $\EuScript{G}$, the user locations in the adjacent segment, inter-segment NLoS fading, blockage states, Willie position in the considered segment and Carol activity in the adjacent segments are fixed. Thus, the corresponding inter-segment interference power is fixed, while the intra-segment Carol-to-Willie fading remains random. Thus, conditioning \eqref{MDP} on $\EuScript{G}$, we obtain \vspace{-0.25em}
\begin{equation}
    \mathbb{P}_{MD,n}|\EuScript{G},\Phi_{C,n}=\Pr\left\{\left\lvert h_{C,n}^{W,n}\right\rvert^2\leq\frac{\Lambda_n-\EuScript{I}_{W,n}}{\EuScript{P}_{C,n}}\bigg|\EuScript{G},\Phi_{C,n}\right\}.
\end{equation}
Now, $\left\lvert h_{C,n}^{W,n}\right\rvert^2$ follows an exponential distribution with mean $\mu_{C,n}^{W,n}=\frac{\Omega_{C,n}^{W,n}}{\left\lVert \Phi_{W,n}-\Phi_{C,n} \right\rVert^{\nu}}$. Hence, the conditional MDP can be expressed as
\vspace{-0.25em}
\begin{equation}
\label{MDP_Conditioned}
    \mathbb{P}_{MD,n}|\EuScript{G},\Phi_{C,n}=
    \begin{cases}
        1-e^{-\frac{\Lambda_n-\EuScript{I}_{W,n}}{\EuScript{P}_{C,n}}\mu_{C,n}^{W,n}}, & \EuScript{I}_{W,n}<\Lambda_n,\\
        0,& \EuScript{I}_{W,n}\geq\Lambda_n.
    \end{cases}
\end{equation}
Finally, averaging \eqref{MDP_Conditioned} over the random network configuration $\EuScript{G}$ yields\vspace{-0.5em}
\begin{equation}
\label{MDP_Final}
    \mathbb{P}_{MD,n}\approx\mathbb{E}_{\EuScript{G},\Phi_{c,n}}\left[\mathbb{P}_{MD,n}|\EuScript{G},\Phi_{C,n}\right].
\end{equation}
Thus, by substituting \eqref{FAP_Final} and \eqref{MDP_Final} into \eqref{DEP}, the semi-analytical DEP can be obtained. Since Willie chooses the detection threshold to minimize the average DEP, the optimal detection threshold is given by $\zeta_n^*=\arg \min_{\zeta_n} \mathbb{P}_{DEP,n}(\zeta_n)$, and the corresponding minimum DEP is $\mathbb{P}_{DEP,n}^*=\mathbb{P}_{DEP,n}\left(\zeta_n^*\right)$.
\vspace{-0.75em}
\subsection{Average Secrecy Capacity Analysis}
The overall ASC of $\EuScript{S}_n$ can be obtained by accounting for Carol's activity probability and can be given as
\begin{equation}
    \mathbb{C}_{sec,n}=(1-p_{act})\mathbb{C}_{sec,n}^{(0)}+p_{act}\mathbb{C}_{sec,n}^{(1)},
\end{equation}
where $\mathbb{C}_{sec,n}^{(q)}= \mathbb{E}_{\EuScript{G}}\left[\mathbb{C}_{sec,n}^{(q)}\mid \EuScript{G}\right]$ and $\mathbb{C}_{sec,n}^{(q)}\mid \EuScript{G}$ denotes the ASC under hypothesis $\EuScript{H}_q$, averaged over the random network configurations. Here, the conditioning on $\EuScript{G}$ fixes the inter-segment interference and Willie's position, while the intra-segment Bob and Carol positions remain random. Their corresponding estimated channel powers are characterized using Lemma \ref{Lemma_1} with the spatial dependence between the legitimate and Willie-side channels retained during the conditional averaging. Thus, for a given network configuration $\EuScript{G}$, the conditional ASC under hypothesis $\EuScript{H}_q$ is obtained from \eqref{Instant_Secrecy_Rate} as
\begin{equation}
\label{57}
    \mathbb{C}_{sec,n}^{(q)}\mid\EuScript{G}=\mathbb{E}\left[\left[\log_2\left(1+\Gamma_{B,n}^{P,q}\right)-\log_2\left(1+\Gamma_{B,n}^{W,q}\right)\right]^{+}\Big| \EuScript{G}\right].
\end{equation}
For $\Gamma_{B,n}^{P,q}>\Gamma_{B,n}^{W,q}$, the instantaneous secrecy rate can equivalently be expressed as\vspace{-0.25em}
\begin{equation}
\label{58}
    \log_2\left(1+\Gamma_{B,n}^{P,q}\right)-\log_2\left(1+\Gamma_{B,n}^{W,q}\right)=\frac{1}{\ln 2}\int_{\Gamma_{B,n}^{W,q}}^{\Gamma_{B,n}^{P,q}}\frac{1}{1+x}dx.
\end{equation}
Hence, averaging \eqref{58} over the remaining random quantities for a given network configuration $\EuScript{G}$ yields
\begin{equation}
\label{59}
\begin{split}
    \mathbb{C}_{sec,n}^{(q)}\mid\EuScript{G}&=\Pr\left\{\Gamma_{B,n}^{P,q}>\Gamma_{B,n}^{W,q}\big|\EuScript{G}\right\}\times\\&\hspace{10mm}\frac{1}{\ln 2}\mathbb{E}\left[\int_{\Gamma_{B,n}^{W,q}}^{\Gamma_{B,n}^{P,q}}\frac{1}{1+x}dx\bigg|\EuScript{G},\Gamma_{B,n}^{P,q}>\Gamma_{B,n}^{W,q}\right].
    \end{split}
\end{equation}
Now, since the integration limits in \eqref{59} are random, for any fixed $x\geq 0$, the term $(1+x)^{-1}$ contributes to the integral only when $x$ lies between the instantaneous wiretap and legitimate SINRs i.e. $\Gamma_{B,n}^{W,q}<x<\Gamma_{B,n}^{P,q}$. Thus, by interchanging the order of expectation and integration, \eqref{59} can be rewritten as
\begin{equation}
\label{QQ}
\begin{split}
    \mathbb{C}_{sec,n}^{(q)}|\EuScript{G}&=\frac{1}{\ln 2}\int_0^{\infty} \frac{\Pr\left\{\Gamma_{B,n}^{W,q}<x, \Gamma_{B,n}^{P,q}>x\Big|\EuScript{G}\right\}}{1+x}dx\\
    &=\frac{1}{\ln 2}\int_0^{\infty} \frac{\EuScript{F}_{\Gamma_{B,n}^{W,q}\Big| \Gamma_{B,n}^{P,q}>x,\EuScript{G}}(x)\left(1-\EuScript{F}_{\Gamma_{B,n}^{P,q}\mid \EuScript{G}}(x)\right)}{1+x}dx.
\end{split}
\end{equation}
Although \eqref{QQ} can be numerically evaluated using composite midpoint quadrature, the accuracy of direct integration may deteriorate at high transmit powers. This is because uniformly spaced quadrature nodes may inadequately capture the variation of the integrand due to the factor $\frac{1}{1+x}$. To mitigate this issue, we introduce the logarithmic transformation which eliminates this factor and effectively provides a denser distribution of quadrature nodes, thereby improving numerical accuracy without introducing additional approximations. Thus, let $y=\ln (1+x)$ which yields $x=e^y-1$ and on differentiating this we get $dy=\frac{dx}{1+x}$. Thus, \eqref{QQ} can be re-written as
\begin{equation}
\begin{split}
    \mathbb{C}_{sec,n}^{(q)}|\EuScript{G}&=\frac{1}{\ln 2}\int_0^{\infty} \EuScript{F}_{\Gamma_{B,n}^{W,q}\Big| \Gamma_{B,n}^{P,q}>e^y-1,\EuScript{G}}(e^y-1)\times\\&\hspace{30mm}\left(1-\EuScript{F}_{\Gamma_{B,n}^{P,q}\mid \EuScript{G}}(e^y-1)\right)dy.
    \end{split}
\end{equation}
\vspace{-0.7em}
\subsubsection{\textbf{Under Hypothesis $\EuScript{H}_0 \text{ (i.e. $q=0$)}$}}
Using \eqref{Gamma_BnP0}, the conditional CDF of $\Gamma_{B,n}^{P,0}$ can be given as
\vspace{-0.5em}
\begin{equation}
\label{XX}
        \EuScript{F}_{\Gamma_{B,n}^{P,0}\mid \EuScript{G}}(x)=\EuScript{F}_{\left\lvert\widetilde{H}_{B,n}^{P,n}\right\rvert^2}\left(\frac{x\left(\EuScript{P}_{B,n}\theta_{e,B,n}^2+\EuScript{I}_{P,n}+\sigma_{P,n}^2\right)}{\EuScript{P}_{B,n}\rho_{B,n}^2}\right).
\end{equation}
Similarly, from \eqref{Gamma_BnW0}, the conditional CDF of Bob's SINR at Willie is obtained as \vspace{-0.75em}
\begin{equation}
\label{YY}
    \EuScript{F}_{\Gamma_{B,n}^{W,0}\mid \EuScript{G},\Phi_{B,n}}(x)=1-e^{-x\frac{\EuScript{I}_{W,n}+\sigma_{W,n}^2}{\EuScript{P}_{B,n}\mu_{B,n}^{W,n}}}
\end{equation}
Substituting \eqref{XX} and \eqref{YY} into \eqref{QQ} yields the conditional ASC under $\EuScript{H}_0$.
\subsubsection{\textbf{Under Hypothesis $\EuScript{H}_1 \text{ (i.e. $q=1$)}$}}
Using \eqref{Gamma_BnP1}, the conditional CDF of $\Gamma_{B,n}^{P,1}$ is given as
\begin{equation}
\label{bhow}
\hspace{-1.25em}\begin{split}
    &\EuScript{F}_{\Gamma_{B,n}^{P,1}\mid \EuScript{G}}(x)\!=\!\!\!\int_0^\infty\!\!\!\EuScript{F}_{\left\lvert\widetilde{H}_{B,n}^{P,n}\right\rvert^2}\left(\frac{x\left(\EuScript{P}_{C,n}\rho_{C,n}^2z\!+\!\EuScript{V}_1\right)}{\EuScript{P}_{B,n}\rho_{B,n}^2}\right)\!f_{\left\lvert\widetilde{H}_{C,n}^{P,n}\right\rvert^2}(z)dz,
\end{split}
\end{equation}
where $\EuScript{V}_1=\EuScript{P}_{B,n}\theta_{e,B,n}^2+\EuScript{P}_{C,n}\theta_{e,C,n}^2+\EuScript{I}_{P,n}+\sigma_{P,n}^2$. Similarly, from \eqref{Gamma_BnW1}, the conditional CDF of Bob's SINR at Willie is obtained as \vspace{-0.65em}
\begin{equation}
\label{meow}
    \EuScript{F}_{\Gamma_{B,n}^{W,1}\mid \EuScript{G},\Phi_{B,n},\Phi_{C,n}}(x)=1-\frac{e^{-\frac{x\left(\EuScript{I}_{W,n}+\sigma_{W,n}^2\right)}{\EuScript{P}_{B,n}\mu_{B,n}^{W,n}}}}{1+\frac{x\EuScript{P}_{C,n}\mu_{C,n}^{W,n}}{\EuScript{P}_{B,n}\mu_{B,n}^{W,n}}}.
\end{equation}
The conditional ASC is subsequently evaluated using the joint probability in \eqref{QQ}. Specifically, the Willie-side fading is averaged using \eqref{YY} and \eqref{meow} while the remaining spatial randomness is accounted for using the channel-power distributions in Lemma \ref{Lemma_1}, thereby preserving the dependence between the legitimate and Willie-side channels.
\vspace{-1em}
\subsection{Covert Rate Analysis}
Since Carol transmits only under $\EuScript{H}_1$, the average covert rate of $\EuScript{S}_n$ accounting for Carol's activity probability, can be given as
\begin{equation}
    \label{64}
    \mathbb{C}_{cov,n}=p_{act}\mathbb{E}_{\EuScript{G}}\left[\mathbb{C}_{cov,n}^{(1)}\mid \EuScript{G}\right],
\end{equation}
where $\mathbb{C}_{cov,n}^{(1)}\mid \EuScript{G}$ denotes the conditional average covert rate for a given network configuration $\EuScript{G}$. Hence, the conditional average covert rate can be written as shown in \eqref{65} (top of next page).
\begin{figure*}
\begin{equation}
\label{65}
\begin{split}
    \mathbb{C}_{cov,n}^{(1)}\mid \EuScript{G}=\int_0^\infty\int_0^\infty \log_2\left(1+\frac{\EuScript{P}_{C,n}\rho_{C,n}^2y}{\Delta_{P,n}\EuScript{P}_{B,n}\rho_{B,n}^2x+\EuScript{P}_{B,n}\theta_{e,B,n}^2+\EuScript{P}_{C,n}\theta_{e,C,n}^2+\EuScript{I}_{P,n}+\sigma_{P,n}^2}\right)f_{\left\lvert\widetilde{H}_{B,n}^{P,n}\right\rvert^2}(x)f_{\left\lvert\widetilde{H}_{C,n}^{P,n}\right\rvert^2}(y)dydx.
\end{split}
\end{equation}
\hrule
 \vspace{-1.25em}
\end{figure*}
The double integral in \eqref{65} is numerically evaluated using the composite midpoint quadrature rule, with the PDF of $\left\lvert\widetilde{H}_{R,n}^{P,n}\right\rvert^2$ given in Lemma \ref{Lemma_1}. The resulting conditional covert rate is subsequently averaged over independent network realizations to obtain the overall covert rate as shown in \eqref{64}.
\vspace{-0.75em}
\section{Results}
\label{Results}
\begin{table}[t]
\centering
\caption{Simulation Parameters.}
\label{tab:PARAMETERS}
\renewcommand{\arraystretch}{0.6}
\begin{tabular}{|c||c|}
 \hline
 \rowcolor{magenta!20} {\bf Parameters} & {\bf Values}\\
 \hline
 \rowcolor{cyan!20}\multicolumn{2}{|c|}{\textbf{General System Parameters}} \\
  \hline
  Overall dimensions $(L\times W)$ (in m) & $200\times 100$\\
  Number of segments $(N)$ & $1 \sim 50$\\
  Waveguide segment height $(d)$ & $5$ m\\
  Attenuation coefficient of waveguide $(\kappa)$ (in dB/m) \cite{11348983} & $0.08\sim0.2$\\
  Effective refractive index of waveguide segment $(\eta_{eff})$ & 1.4\\
  Background noise power $(\sigma_{X,n}^2)$ & $-70$ dBm\\
  Considered segment index $(n)$ & $1< n< N$\\
  Imperfect SIC parameter $(\Delta_{P,n})$ (ipSIC, pSIC) & 0.1, 0\\
  Number of pilot symbols $(\EuScript{N}_{R,n})$ (ipCSI, pCSI) & 10, $10^6$\\
  Pilot estimation degradation $(\delta_{R,n})$ (non-ideal, ideal) & 100, 1\\
  Number of Monte Carlo simulations & $10^6$\\
  Composite midpoint quadrature points & 30\\
  \hline 
  \rowcolor{cyan!20}\multicolumn{2}{|c|}{\textbf{Channel Parameters}} \\
  \hline
  Carrier Frequency $(f_{C})$ & 28 GHz\\
  Blockage parameter $\left(\phi_{R,i}^{X,n}\right)$ & $0.0002$\\
  Average NLoS power $(\Omega_{R,i}^{W,n})$ \cite{wang2026performance} & $-50$ dBm\\
  Path loss factor $(\nu)$ \cite{TCOM} & 6\\
  \hline 
  \rowcolor{cyan!20}\multicolumn{2}{|c|}{\textbf{Parameters for Bob (Secrecy User)}} \\
  \hline
  Bob Transmit power of $\EuScript{S}_n$ $(\EuScript{P}_{B,n})$ & 10-70 dBm\\
  Bob power from adjacent segment $(\EuScript{P}_{B,n+1}/\EuScript{P}_{B,n-1})$& 10 dBm\\
  Target rate for Bob $(\EuScript{R}_{sec,Th})$ & 0.5 bps/Hz\\
  \hline
  \rowcolor{cyan!20}\multicolumn{2}{|c|}{\textbf{Parameters for Carol (Covert User)}} \\
  \hline
  Carol power of $\EuScript{S}_n$ $(\EuScript{P}_{C,n})$ & 0-36 dBm\\
  Carol power from adjacent segment $(\EuScript{P}_{C,n+1}/\EuScript{P}_{C,n-1})$& 5 dBm\\
  Normalised Detection Threshold $\left(\widehat{\zeta}_n\right)$ (in dB) & $-2$ to $2$\\
  Activity probability $\left(p_{act}=p_{act,n}=p_{act,n+1}=p_{act,n-1}\right)$ & $0.5$\\
  \hline
\end{tabular}
\end{table}
In this section, the secrecy and covert communication performance of the proposed framework is evaluated to demonstrate its operational behaviour and effectiveness. The simulation results are presented and compared with the corresponding semi-analytical results to validate the accuracy of the developed analysis. It can be observed that the simulation results closely match the analytical results, represented using \say{Ana} in the figures, across all the considered cases, thereby verifying the accuracy of the derived expressions for SOP, ASC, DEP, and CR. The proposed framework is denoted by EDMA-NOMA-SC, where the secrecy user and the intermittently active covert user share the same resource through NOMA within each segment. It is compared with the following baseline schemes
\begin{itemize}
    \item \textbf{EDMA-OMA:} The EDMA architecture is retained, while the secrecy and covert users are served over orthogonal time resources. Accordingly, within each segment, the secrecy and covert users are each allocated half of the available transmission interval. The covert user remains intermittently active and transmits in its allocated interval with probability $p_{act}$. In the figures, the corresponding secrecy and covert results are denoted by EDMA-OMA-S and EDMA-OMA-C, respectively.
    \item \textbf{Conv. NOMA-PASS:} The conventional NOMA-PASS baseline is obtained by setting $N=1$, such that the system reduces to a single-segment PASS configuration for the entire service area. Consequently, no adjacent segments are present and the corresponding inter-segment interference vanishes. Within this single segment, the secrecy and covert users simultaneously share the same resource through NOMA and the covert user transmits intermittently with probability $p_{act}$. This scheme is denoted by Conv. NOMA-PASS-SC in the figures.
    \item \textbf{Conv. OMA-PASS:} The conventional OMA-PASS system is also obtained by setting $N=1$ as discussed above. The secrecy and covert users are then assigned orthogonal transmission intervals, with each user occupying half of the available time resource. The covert user transmits within its assigned interval with probability $p_{act}$. The corresponding secrecy and covert results are denoted by Conv. OMA-PASS-S and Conv. OMA-PASS-C, respectively, in the figures.
\end{itemize}
For the OMA-based frameworks, the secrecy and covert users are each allocated half of the available transmission interval. Accordingly, the corresponding secrecy and covert rates are multiplied by $\frac{1}{2}$. Equivalently, for the SOP evaluation, the secrecy-rate threshold is scaled to $2\EuScript{R}_{sec,Th}$. However, the DEP detection threshold remains unchanged since it is determined from Willie’s received energy during the covert-user transmission interval. The system parameters used throughout the evaluation are summarized in Table \ref{tab:PARAMETERS}.

\vspace{-0.5em}
\subsection{Secrecy Performance Comparison}
\begin{figure}[t]
\vspace{-0.5em}
\centering
    \includegraphics[scale=0.5]{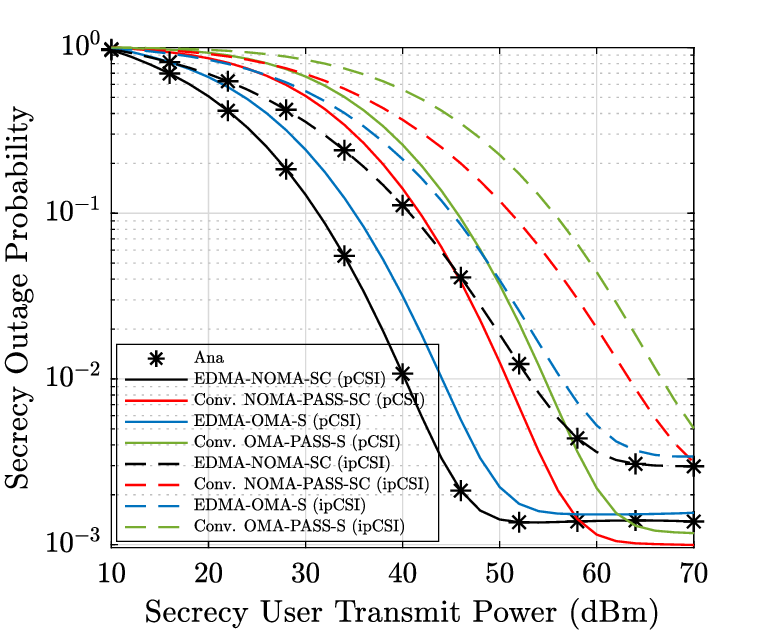}
    \caption{SOP wrt $\EuScript{P}_{B,n}$ under pCSI and ipCSI for the considered frameworks.}
    \label{SOP_Power}
   
\end{figure}
\par \figurename{~\ref{SOP_Power}} illustrates the SOP versus the secrecy user's transmit power, $\EuScript{P}_{B,n}$, for the considered frameworks under pCSI and ipCSI scenarios. Here we consider $N$=3 for the EDMA-based frameworks. As $\EuScript{P}_{B,n}$ increases, the SOP initially decreases for all schemes owing to the increased received power of Bob's transmission at Alice, while the degradation caused by ipCSI becomes evident through the higher SOP compared with the corresponding pCSI cases. In the low-to-moderate power region, the EDMA-based schemes generally outperform Conv. NOMA-PASS-SC and Conv. OMA-PASS-S, since segmentation reduces both the propagation distance between Bob and its associated PA and the in-waveguide propagation loss. Moreover, the NOMA-based schemes outperform their OMA counterparts over this region, as NOMA allows Bob to utilize the entire transmission interval, whereas the OMA-based schemes allocate only half of the available resources to the secrecy transmission. However, the SOP curves gradually approach an outage floor at high transmit powers because the SINRs of Bob at both Alice and Willie increase with $\EuScript{P}_{B,n}$ causing the corresponding secrecy-rate difference to gradually saturate. In this region, the conventional PASS schemes can achieve lower SOP than their EDMA counterparts since they are free from inter-segment interference and allow Bob and Willie to be distributed over the entire service region, potentially resulting in a larger Bob-Willie separation than when both are confined within an EDMA segment. Consequently, as the benefit of increasing $\EuScript{P}_{B,n}$ diminishes, the weaker eavesdropping link and absence of inter-segment interference can favour the conventional PASS schemes. 
\begin{figure}[t]
\centering
    \includegraphics[scale=0.5]{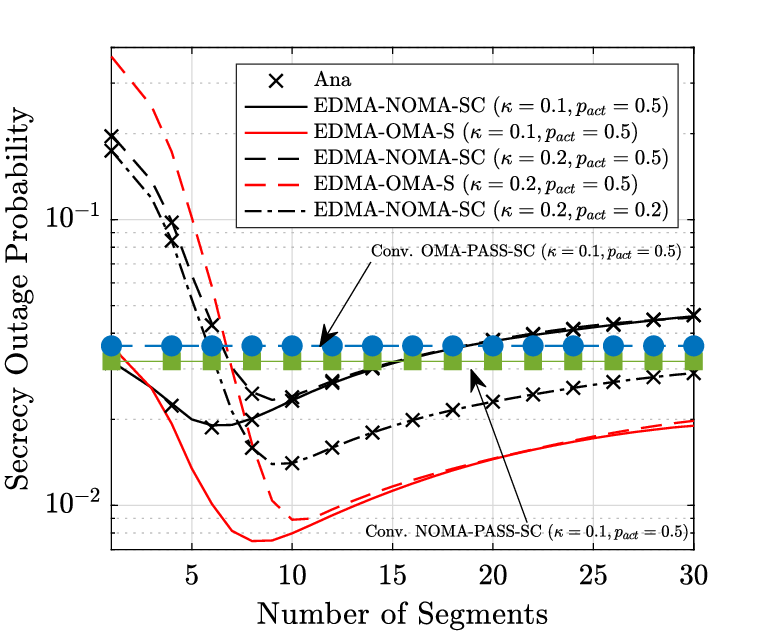}
    \caption{SOP wrt $N$ for varying $\kappa$ and $p_{act}$ under ipCSI scenario for $L\times W=125$ m$\times 75$ m.}
    \label{SOP_Segments}
\end{figure}
\par \figurename{~\ref{SOP_Segments}} illustrates the SOP with respect to $N$ for different values of $\kappa$ and $p_{act}$. Here, we consider $\EuScript{P}_{B,n}=\EuScript{P}_{B,n-1}=\EuScript{P}_{B,n+1}=40$ dBm and $\EuScript{P}_{C,n}=\EuScript{P}_{C,n-1}=\EuScript{P}_{C,n+1}=30$ dBm. For EDMA-NOMA-SC, increasing $N$ initially reduces the SOP because shorter segments decrease both the in-waveguide propagation distance and the Bob-to-PA distance. Beyond an intermediate $N$, however, inter-segment interference becomes increasingly dominant and outweighs these gains, causing the SOP to increase. A similar trend is observed for EDMA-OMA-S. At small $N$, EDMA-OMA-S performs worse because Bob is allocated only half of the transmission interval, resulting in a higher effective secrecy-rate threshold. As $N$ increases, this penalty becomes less dominant, while the absence of intra-segment interference from Carol makes EDMA-OMA-S increasingly competitive and eventually better than its NOMA counterpart. Increasing $\kappa$ degrades secrecy performance, particularly for small $N$, whereas its impact diminishes for larger $N$ due to the reduced PA-to-feed distance. A lower $p_{act}$ also improves the NOMA secrecy performance by reducing both intra- and inter-segment interference from Carol. To summarise, the results indicate the presence of an optimal value of $N$. Finally, it should be noted that excessive segmentation can erode the secrecy gains of EDMA owing to increased inter-segment interference and may even make its performance worse than the corresponding conventional PASS schemes.
\begin{figure}[t]
\centering
    \includegraphics[scale=0.5]{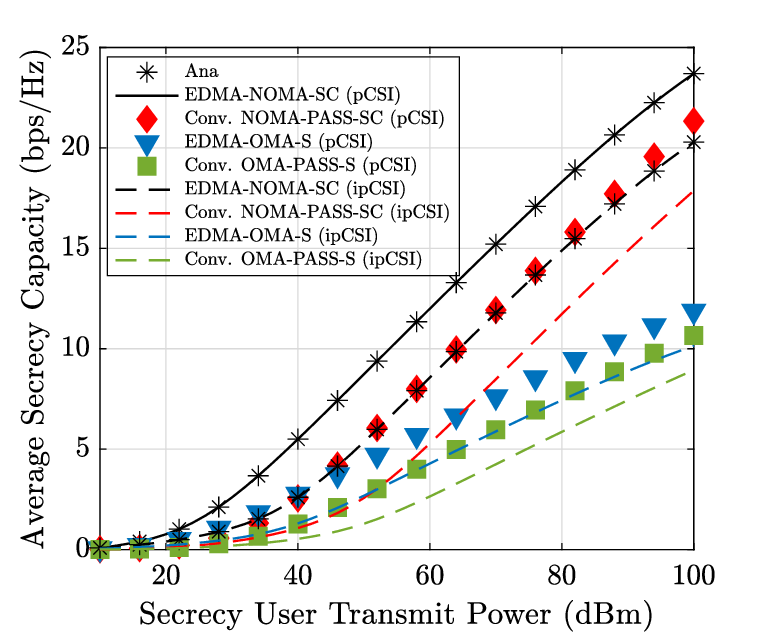}
    \caption{ASC wrt $\EuScript{P}_{B,n}$ under pCSI and ipCSI for the considered frameworks.}
    \label{Secrecy_Rate}
\end{figure}
\par \figurename{~\ref{Secrecy_Rate}} illustrates the ASC performance of the considered frameworks wrt $\EuScript{P}_{B,n}$ under pCSI and ipCSI scenarios for $N$=3. It can be observed that as $\EuScript{P}_{B,n}$ increases, the ASC increases for all schemes owing to the improved received power of Bob's transmission at Alice. The proposed EDMA-NOMA-SC framework exhibits the highest ASC performance over the considered power range owing to the reduced Bob-to-PA propagation distance and the associated in-waveguide propagation loss. The conventional NOMA-PASS-SC scheme achieves the second highest ASC, although its performance is lower than the proposed EDMA-NOMA-SC due to the larger propagation distances associated with the conventional architecture. Moreover, the OMA-based schemes exhibit lower ASC than their corresponding NOMA-based counterparts because only half of the available transmission interval is allocated to the secrecy user. However, EDMA-OMA-S still outperforms Conv. OMA-PASS-S owing to the reduced in-waveguide attenuation and Bob-to-PA propagation distance. Furthermore, it can be observed that the ASC performance degrades under the ipCSI scenario for all the frameworks as channel-estimation errors reduce the effective SINR at Alice and hence reduce the ASC. Although the ASC continues to increase over the considered transmit-power range, its growth gradually becomes slower at higher powers because the achievable rates at both Alice and Willie increase with $\EuScript{P}_{B,n}$, thereby reducing the ASC gain.
\vspace{-0.5em}
\subsection{Covert Communication Performance Comparison}
In this sub-section, we illustrate the performance of the proposed EDMA-NOMA-SC framework in comparison with the benchmark schemes in terms of covert communication performance. From the covertness perspective, higher FAP and MDP indicate larger probabilities of incorrect decisions by Willie under $\EuScript{H}_0$ and $\EuScript{H}_1$, respectively, while a higher DEP represents greater overall detection uncertainty and hence better covertness performance.
\begin{figure}[t]
\centering
    \includegraphics[scale=0.5]{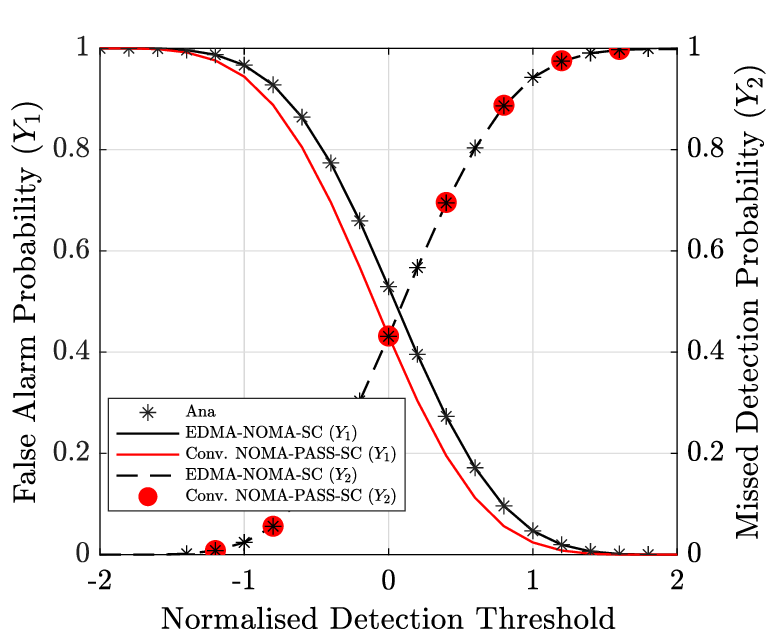}
    \caption{FAP ($Y_1$) and MDP ($Y_2$) performance comparison wrt normalised detection threshold.}
    \label{FAP_MDP}
\end{figure}
\par The performance comparison in terms of FAP ($Y_1$) and MDP ($Y_2$) wrt the normalized detection threshold for the considered EDMA-NOMA-SC and Conv. NOMA-PASS-SC frameworks is illustrated in \figurename{~\ref{FAP_MDP}}. The OMA-based frameworks are not separately shown because, for a conservative analysis, Willie is assumed to possess pCSI and perfectly cancel Bob's secrecy signal through SIC before detecting Carol's signal. Consequently, the post-SIC observation used for covert detection in the NOMA case is identical to that of the corresponding OMA case under the same system parameters, resulting in the same FAP and MDP. Furthermore, the normalised detection threshold, $\widehat{\zeta}_n$, is varied from $-2$ to $2$ dB, while the corresponding linear detection threshold is obtained as $\zeta_n=10^{-10}10^{\widehat{\zeta}_n/10}$. Now, it can be observed that as the detection threshold increases, the FAP decreases because Willie decides in favour of $\EuScript{H}_1$ only when the received energy exceeds the detection threshold. Hence, under $\EuScript{H}_0$, increasing the threshold reduces the probability that the received energy crosses this threshold, thereby lowering the FAP. On the other hand, the MDP increases because, under $\EuScript{H}_1$, a larger threshold increases the probability that the received energy remains below the threshold, causing Willie to incorrectly decide in favour of $\EuScript{H}_0$ even when Carol is active. Moreover, EDMA-NOMA-SC exhibits a higher FAP than Conv. NOMA-PASS-SC because the inter-segment interference contributes to Willie's received energy under $\EuScript{H}_0$, thereby increasing the probability that the received energy exceeds the detection threshold. In contrast, the MDP curves of the two frameworks remain nearly identical. This is because, under $\EuScript{H}_1$, Carol's covert signal is additionally present at Willie and provides the dominant contribution to the received energy relevant to the missed-detection event, making the effect of the additional inter-segment interference comparatively smaller.
\begin{figure}[t]
\centering
    \includegraphics[scale=0.5]{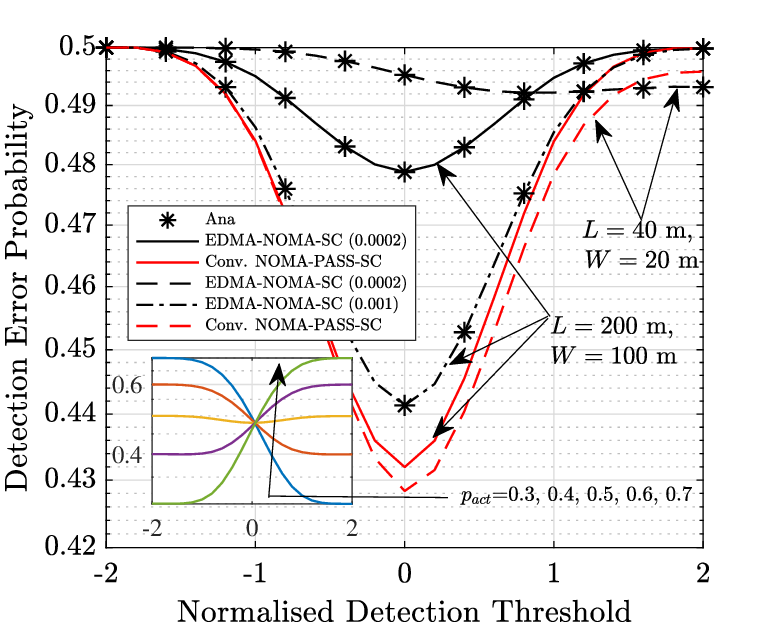}
    \caption{DEP wrt $\widehat{\zeta}_n$ for varying $L\times W$, $\phi_{R,i}^{X,n}$, and $p_{act}$.}
    \label{DEP_Threshold}
\end{figure}
\par \figurename{~\ref{DEP_Threshold}} illustrates the performance of DEP wrt $\widehat{\zeta}_n$ for the considered EDMA-NOMA-SC and Conv. NOMA-PASS-SC frameworks. As observed from \figurename{~\ref{FAP_MDP}}, increasing $\widehat{\zeta}_n$ decreases the FAP while increasing the MDP. Consequently, since DEP is also a function of FAP and MDP, it initially decreases as the reduction in FAP dominates, reaches a minimum at an intermediate threshold, and subsequently increases when the contribution of MDP becomes dominant. The minimum point therefore corresponds to the detection threshold most favourable to Willie, since Willie selects the threshold that minimizes the DEP. For the service area of $L=$200 m and $W$=100 m, the proposed EDMA-NOMA-SC framework exhibits a noticeably higher DEP than Conv. NOMA-PASS-SC over the considered threshold range. Unlike the conventional framework, the EDMA framework experiences inter-segment interference at Willie, which increases the uncertainty in distinguishing between $\EuScript{H}_0$ and $\EuScript{H}_1$, thereby improving the covertness performance of the proposed EDMA-NOMA-SC framework. Furthermore, when the blockage parameter is increased from $0.0002$ to $0.001$, the probability of the inter-segment links being available decreases, thereby reducing the corresponding interference at Willie. This reduction in interference lowers the DEP and hence degrades the covertness performance, further highlighting the beneficial role of inter-segment interference in the considered framework. 
\par Now, when the service area is reduced to $L=$40 m and $W=$20 m, owing to the smaller dimensions, Carol and Willie are, on average, located closer to each other, which strengthens the Carol-to-Willie link. In the low-threshold region, however, the DEP is mainly governed by the FAP. Hence, the increase in received energy caused by the stronger inter-segment interference under $\EuScript{H}_0$ has the more pronounced effect, leading to a higher DEP of the proposed EDMA-NOMA-SC framework when compared to the Conv. NOMA-PASS-SC framework as can be observed in \figurename{~\ref{DEP_Threshold}}. As the detection threshold increases, the MDP contribution becomes dominant. In this region, the stronger Carol-to-Willie link becomes more influential, since Carol’s transmission under $\EuScript{H}_1$ is received with higher power and is therefore less likely to be missed. Consequently, the beneficial effect of inter-segment interference on covertness is outweighed by the stronger covert-signal component, causing EDMA-NOMA-SC to exhibit a lower DEP than Conv. NOMA-PASS-SC in the high-threshold region. Finally, the inset illustrates the impact of Carol’s activity probability, $p_{act}$ on the DEP. At low detection thresholds, the FAP contribution dominates. Hence, a lower $p_{act}$  assigns a larger weight $(1-p_{act})$ to the FAP term, resulting in a higher DEP. Therefore, $p_{act}$=0.3 yields the highest DEP in this region, while $p_{act}$=0.7 yields the lowest. In contrast, at high detection thresholds, the MDP contribution becomes dominant. Hence, a higher $p_{act}$ assigns a larger weight to the MDP term, causing the ordering of the curves to reverse. Around the intermediate-threshold region, the FAP and MDP contributions become comparable, causing the corresponding DEP curves to converge.
\begin{figure}[t]
\centering
    \includegraphics[scale=0.5]{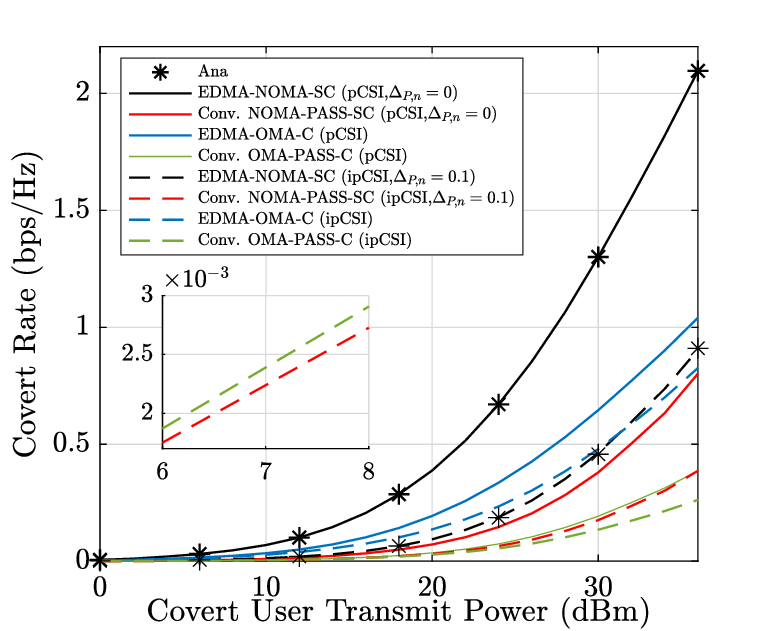}
    \caption{CR performance comparison wrt $\EuScript{P}_{C,n}$ under ideal and practical scenarios for the considered frameworks.}
    \label{Covert_Rate}
\end{figure}
\par \figurename{~\ref{Covert_Rate}} illustrates the CR performance of the considered frameworks wrt $\EuScript{P}_{C,n}$ under ideal and practical scenarios. Here, the ideal scenario refers to the case with pCSI and pSIC, whereas the practical scenario considers ipCSI and ipSIC. It can be observed that the CR increases with $\EuScript{P}_{C,n}$ for all the schemes, since increasing Carol's transmit power improves the received signal strength at Alice and consequently increases the achievable rate of the covert link. Under the ideal scenario, EDMA-NOMA-SC achieves the highest CR due to the reduced Carol-to-PA propagation distance and in-waveguide propagation loss. EDMA-OMA-C achieves the second-highest CR, as the reduced in-waveguide attenuation provided by segmentation and the enhanced channel gain due to the reduced distance between the nodes compensate for the half resource allocation of OMA, while Conv. NOMA-PASS-SC follows owing to its comparatively larger propagation distance. Under the practical scenario, the CR decreases due to ipCSI, while the NOMA-based schemes are additionally affected by residual Bob interference under ipSIC. However, increasing $\EuScript{P}_{C,n}$ also improves Carol's channel-estimation quality by increasing the pilot SNR and $\rho_{C,n}$, thereby reducing the estimation-error variance. Consequently, the impact of ipCSI becomes less pronounced, while the increasing desired signal power reduces the relative impact of the residual interference caused by ipSIC at higher $\EuScript{P}_{C,n}$. The practical EDMA-NOMA-SC and EDMA-OMA-C curves therefore remain close over part of the power range, since the full-resource advantage of NOMA is partly offset by residual SIC interference. At higher $\EuScript{P}_{C,n}$, the stronger desired signal reduces the relative impact of this residual interference, allowing NOMA to regain its advantage. A similar trend is observed for the conventional schemes in the inset, where Conv. OMA-PASS-C initially outperforms Conv. NOMA-PASS-SC, but the ordering reverses as $\EuScript{P}_{C,n}$ increases.

\balance
\section{Conclusion}
\label{Conclusion}
This paper proposed an EDMA-NOMA-SC framework with SWAN-enabled PASS for jointly supporting uplink secrecy and covert communications. In each segment, a secrecy user and an intermittently active covert user share the same transmission resource through NOMA, while segmentation reduces the user-to-PA propagation distance and in-waveguide attenuation. To capture practical operating conditions, distance-dependent blockage, inter-segment interference, ipCSI, and ipSIC were considered. Semi-analytical expressions were developed to characterize the SOP, ASC, FAP, MDP, DEP, and CR, and their accuracy was verified through simulations. The proposed framework generally achieved improved secrecy performance, although excessive segmentation can degrade performance when inter-segment interference outweighs the propagation benefits. The results also showed the adverse effects of ipCSI and ipSIC, while improved pilot-SNR-based channel estimation reduces the impact of ipCSI at higher transmit powers. From the covertness perspective, inter-segment interference can increase Willie’s detection uncertainty and improve the DEP, with its impact depending on the detection threshold, service-area dimensions, blockage conditions, and covert-user activity probability.
\bibliographystyle{IEEEtran}
\bibliography{Bibliography.bib}
\end{document}